\documentclass[12pt,epsfig]{article}

\usepackage{graphicx}
\usepackage{subfig}
\usepackage{amsfonts}
\usepackage{amssymb}
\usepackage{amsmath}
\usepackage{mathdots}
\usepackage{placeins}
\usepackage{bbold}

\newskip\humongous \humongous=0pt plus 1000pt minus 1000pt
  \newif\ifdtup

\let\l\left
\let\r\right
\let\de\partial
\let\mrm\mathrm
\let\mcal\mathcal

\let\mbb\mathbb

\let\dag\dagger

\def\frac#1#2{ {{#1} \over {#2} }}

\def\eg{\hbox{\em e.g. }}
\def\ie{\hbox{\em i.e. }}

\def\beq{\begin{equation}}
\def\eeq{\end{equation}}

\newcommand{\beqa}{\begin{eqnarray}}
\newcommand{\eeqa}{\end{eqnarray}}
\newcommand{\ba}{\begin{array}}
\newcommand{\ea}{\end{array}}
\newcommand{\bmat}{\begin{pmatrix}}
\newcommand{\emat}{\end{pmatrix}}
\newcommand{\bcas}{\begin{cases}}
\newcommand{\ecas}{\end{cases}}
\newcommand{\muh}{{\hat{\mu}}}
\newcommand{\nuh}{{\hat{\nu}}}

\newcommand{\dd}{{\mathrm{d}}}
\newcommand{\DD}{{\mathcal{D}}}
\newcommand{\tr}{{\mathrm{Tr}}}
\newcommand{\id}{{\mathbb{1}}}

\def\de{\delta}

\let\de\partial
\let\mrm\mathrm
\let\mcal\mathcal

\let\mbb\mathbb
\let\dag\dagger

\begin{document}

\title
{One-dimensional QCD in thimble regularization.}

\author
{F.~Di~Renzo and G.~Eruzzi\thanks{Present address: Maps spa, I-43122
    Parma, Italy}\\
\small{Dipartimento di Scienze Matematiche, Fisiche e Informatiche, Universit\`a di Parma} \\
\small{and INFN, Gruppo Collegato di Parma} \\
\small{I-43100 Parma, Italy}\\
}

\maketitle

\begin{abstract}
QCD in 0+1 dimensions is numerically solved via thimble
regularization. In the context of this toy model, a general 
formalism is presented for $SU(N)$ theories. The sign problem 
that the theory displays is a genuine one, stemming from a (quark) 
chemical potential. Three stationary points are present in the 
original (real) domain of integration, so that contributions 
from all the thimbles associated to them are to be taken into 
account: we show how semi-classical computations can provide hints 
on the regions of parameter space where this is absolutely crucial.
Known analytical results for the chiral condensate and the Polyakov 
loop are correctly reproduced: this is in particular trivial at 
high values of the number of flavors $N_f$. In this regime we notice
that the single thimble dominance scenario takes place (the dominant
thimble is the one associated to the identity). At low values of $N_f$ 
computations can be more difficult. It is important to stress that 
this is not at all a consequence of the {\em original} sign problem 
(not even via the residual phase). The latter is always 
under control, while accidental, delicate cancelations of 
contributions coming from different thimbles can be in place in 
(restricted) regions of the parameter space.      
\end{abstract}

\section{Introduction}

The very first proposal of thimble regularization was intended to 
extend our capabilities to properly define quantum field 
theories \cite{Witten}. After a while it was realized that it was a 
very natural and powerful candidate solution for the sign problem 
\cite{AUthimble, Kiku}. Since then it has attracted quite a lot of 
attention. By now also other proposals have been put forward 
which are more or less inspired by the thimble approach, \eg the 
holomorphic gradient flow \cite{PaulAndrei} or the idea of 
combining the latter with the complex Langevin method \cite{LangeThimbles}.

All in all, the basic idea underlying thimble regularization 
amounts to deforming the original domain of integration of a given
field theory into a new one, which is made of one or more manifolds. 
These manifolds live in the complexification of the original domain 
of integration; in the original formulation (which is the one we adhere
to) they are the Lefschetz Thimbles themselves. Thimbles are defined 
as the union of the Steepest Ascent (SA) paths attached to critical 
points $\sigma$ of the (complexified) action and have the same real 
dimension of the original manifold.  
Proceeding straight to the field theoretic quantities one is
interested in, we denote the thimbles ${\mathcal{J}_{\sigma}}$ 
and in a sketchy way we write 
\begin{equation}
\label{eq:basicO}
\langle O \rangle = \frac{\sum_{\sigma} n_{\sigma} \,
  e^{-i\,S_I\left(p_{\sigma}\right)} \int_{\mathcal{J}_{\sigma}}
  \mathrm{d}z\, e^{-S_R} \,O\, e^{i\omega}}{\sum_{\sigma} n_{\sigma} \,
  e^{-i\,S_I\left(p_{\sigma}\right)} \int_{\mathcal{J}_{\sigma}}
  \mathrm{d}z\, e^{-S_R}  \, e^{i\omega}}
\end{equation}
where the $z$ are a shortcut for {\em complexified} field configurations 
and the $p_{\sigma}$ stand for the configurations which are the 
stationary points of the action S, the sum formally 
extending to all of them, even though the $n_{\sigma}$ can be zero
(thus, not all the critical points do contribute). The action is written 
in terms of a real part $S_R$ and of an imaginary part $S_I$. In the previous 
formula the denominator reconstructs the partition function $Z$.
Notice that a positive measure $e^{-S_R}$ is in place and constant
phases $e^{-i\,S_I\left(p_{\sigma}\right)}$ have been factored out of
the integrals. This is a consequence of the main virtue 
the thimbles have: the imaginary part of the action stays
constant on them. 
A so-called {\em residual phase} $e^{i \omega}$ is there that 
accounts for the relative orientation
between the canonical complex volume form and the real
volume form, characterizing the tangent space of the
thimble. 

Solving the sign problem via a deformation of the integration 
domain is conceptually satisfying and the thimble approach is
potentially very powerful. However one can not omit difficulties: 
thimbles are non-trivial manifolds, for which a local characterisation 
is missing and thus, not surprisingly, devising Monte Carlo methods to 
sample integrals on thimbles is a delicate issue. Moreover, recent 
works have stressed how taking into account multiple thimbles can be 
tricky \cite{PaulAndrei,YuyaMULTI,KikuMULTI}. Finally, the final goal of
virtually any attempt to solve the sign problem is to eventually
attack the study of the QCD phase diagram, and so in the end one
struggles to tackle that ultimate goal. 

Despite its simplicity, the study at hand addresses virtually all the
issues we have just sketched. We will present a numerical study of 
QCD in $0+1$ dimensions, showing that thimble regularization can solve 
it: analytical results are known ({\em e.g.} for the chiral condensate 
and the Polyakov loop) and those have been obtained via Monte Carlo
simulations on thimbles, on a wide range of values for the number of
flavor $N_f$, the mass parameter $m$ and the ratio 
$\mu/T$ (results depend on the chemical potential $\mu$ via this
ratio; $T$ is the temperature). 
Even if in the end there is no real gauge symmetry in place, the model is a 
perfect ground to see the thimble formalism for $SU(N)$ theory 
at work. Moreover, the sign problem one has to tackle is a genuine one,
originating from the (quark) chemical potential via the fermionic
determinant. We have already presented preliminary results of this
study in \cite{LAT16}; for other, independent work on this subject 
see also \cite{ChrLAT16}.  

The paper is organized as follows: in section 2 we discuss $0+1$ QCD
and how to treat it in thimble regularization, in particular
enlightening the role of symmetry and discussing the semiclassical
approximation; in section 3 we present our results, both by flat,
crude Monte Carlo and by importance sampling in the steepest ascents
space; finally, in section 4 we present our conclusions.

\section{Thimble regularization for $0+1$-dimensional QCD} 

\subsection{QCD in $0+1$ dimensions} 
We shall study a lattice formulation of Quantum Chromodynamics in
$0+1$ dimensions. There is some abuse of terminology in naming 
the theory at hand QCD. There can be no Yang-Mills action (and no 
plaquette either) in less than two dimensions. Nevertheless, though much 
simpler than its $4$-dimensional counterpart, this model provides 
an excellent setting to test the thimble formalism for $SU(N)$
theories. Moreover, the sign problem is the genuine one as in real 
QCD, being due to the presence of a (quark) chemical potential. 
We have \emph{staggered} fermions on a one-dimensional lattice 
with (even) $N_t$ sites in the temporal direction. The lattice extent 
is related to the temperature by $aN_t=1/T$, where $a$ is the lattice 
spacing. The partition function of the theory for $N_f$ degenerate 
quark flavours of mass $m$ is

\begin{equation*}
Z_{N_f}=\int\prod\limits_{i=1}^{N_t}\dd U_i\,{\det}^{N_f}(aD)
\end{equation*}

where $D$ is the lattice staggered Dirac operator

\begin{equation*}
(aD)_{ii'}=am\delta_{ii'}+\frac{1}{2}\l(e^{a\mu}U_i\tilde{\delta}_{i',i+1}-e^{-a\mu}U^\dag_{i-1}\tilde{\delta}_{i',i-1}\r)
\end{equation*}

and $\tilde{\delta}_{ii'}$ is the anti-periodic Kronecker delta. 
Explicitly

\begin{equation*}
aD=
\begin{pmatrix}
am & e^{a\mu}U_1/2 & 0 & \cdots & 0 & e^{-a\mu}U^\dag_{N_t}/2 \\
-e^{-a\mu}U^\dag_1/2 & am & e^{a\mu}U_2/2 & \cdots & 0 & 0 \\
\vdots & \vdots & \vdots & \ddots & \vdots & \vdots \\
0 & 0 & 0 & \cdots & am & e^{a\mu}U_{N_t-1}/2 \\
-e^{a\mu}U_{N_t}/2 & 0 & 0 & \cdots & -e^{-a\mu}U^\dag_{N_t-1}/2 & am
\end{pmatrix}
\end{equation*}

By an appropriate gauge transformation we can set to $\id$ all the links $\{U_i\}$ except one. 
The only remaining link is simply the Polyakov loop $U\equiv U_{N_t}$,
so that we are effectively left with an $\mrm{SU}(3)$ one-link model. 
We have

\begin{equation*}
\det(aD)=\det
\begin{pmatrix}
am & e^{a\mu}/2 & 0 & \cdots & 0 & e^{-a\mu}U^\dag/2 \\
-e^{-a\mu}/2 & am & e^{a\mu}/2 & \cdots & 0 & 0 \\
\vdots & \vdots & \vdots & \ddots & \vdots & \vdots \\
0 & 0 & 0 & \cdots & am & e^{a\mu}/2 \\
-e^{a\mu}U/2 & 0 & 0 & \cdots & -e^{-a\mu}/2 & am
\end{pmatrix}
\end{equation*}

When $N_t$ is even the Dirac determinant can be shown to be 
equal to the determinant of a $3\times3$ matrix. Namely, 
the partition function we have to compute is 

\begin{equation}
Z_{N_f}=\int\limits_{\mrm{SU}(3)}\dd U\,{\det}^{N_f}\l(A\,\id_{3\times 3}+e^{\mu/T}U+e^{-\mu/T}U^\dag\r)\label{eq:qcd01_Zint}
\end{equation}

where $A=2\cosh(\mu_c/T)$ and $\mu_c=\sinh^{-1}(m)$ (from now on, we 
set $a=1$ in all the calculations). As usual, the quark determinant 
can be turned into an effective action

\begin{equation*}
Z_{N_f}=\int\limits_{\mrm{SU}(3)}\dd U\,e^{-S(U)}
\end{equation*}

with

\begin{equation*}
S(U)=-N_f\tr\log M(U) = -N_f\tr\log \left(A\,\id_{3\times 3}+e^{\mu/T}U+e^{-\mu/T}U^{-1}\right).
\end{equation*}

We will be concerned with three main observables. The chiral
condensate is the first one 
\begin{equation*}
\Sigma\equiv T\frac{\de}{\de m}\log Z=T\l\langle N_f\tr\l(M^{-1}\frac{\de M}{\de m}\r)\r\rangle=N_f\sqrt{\frac{A^2-4}{m^2+1}}\l\langle\tr\l(M^{-1}\r)\r\rangle.
\end{equation*}
The other two are the Polyakov loop $\langle\tr\,U\rangle$ and the
anti-Polyakov loop
$\langle\tr\,U^\dag\rangle=\langle\tr\,U\rangle_{\mu\rightarrow-\mu}$. The
latter two can be related to the quark number density $n\equiv
T\frac{\de}{\de\mu}\log Z$ by a relation which takes quite different
forms for different values of $N_f$ \cite{qcd01_jacques}. 
We will have numbers to compare to, since 
analytical results for $0+1$ QCD are available
\cite{qcd01_jacques,qcd01_bilic,qcd01_ravagli,qcd01_jacques_pos}.
We also notice that the
sign problem has also been solved by means of the so called
\emph{subset method} \cite{qcd01_jacques} and complex 
Langevin \cite{GertKim}. 
 
In the following we will be interested in critical points, \ie
stationary points of the action: to each of
them a thimble will be attached. In order to write the equations of
motion we first of all introduce the Lie derivative
\begin{equation}
\label{eq:nable}
\nabla^af\left(U\right)=\lim_{\alpha\rightarrow 0}\frac{1}{\alpha}\left[f\left(e^{i\alpha T^a}U\right)-f\left(U\right)\right]=\frac{\delta}{\delta\alpha}f\left(e^{i\alpha T^a}U\right)\biggl|_{\alpha=0}
\end{equation}

Stationary points are now defined as solution of 
$\nabla S(U)=\sum_a T^a\nabla^a S(U)=0$
where 
\begin{equation*}
\nabla^a S(U)=-i\,N_f\tr\l[M^{-1}(U)\l(e^{\mu/T}T^aU-e^{-\mu/T}U^{-1}T^a\r)\r].
\end{equation*}

We are also interested in the Hessian
\begin{align*}
&\nabla^b\nabla^a S(U)=N_f\tr\biggl[M^{-1}(U)\bigl[\bigl(e^{\mu/T}T^aT^bU+e^{-\mu/T}U^{-1}T^bT^a\bigr)\\
&-\bigl(e^{\mu/T}T^bU-e^{-\mu/T}U^{-1}T^b\bigr)M^{-1}(U)\bigl(e^{\mu/T}T^aU-e^{-\mu/T}U^{-1}T^a\bigr)\bigr]\biggr]
\end{align*}

There are three critical points $\{U_k=e^{2\pi i k/3}\id\}$ with
$k=0,1,2$.

\subsection{Thimble regularization for $SU(N)$ theory}

The basic ingredient we need to construct thimbles for $SU(N)$ theory 
were already discussed in \cite{AUthimble} and more recently in 
\cite{LAT15}. Here we recollect the results we need.

\noindent  
First of all, we need to 
complexify the degrees of freedom. Going to complex fields means
$$
\mrm{SU}\l(N\r)\ni U=e^{i x_aT^a}\rightarrow e^{i z_aT^a}=e^{i \l(x_a+i y_a\r)T^a}\in\mrm{SL}\l(N,\mbb{C}\r).
$$
We want to stress that 
$$
\mathrm{SU}\left(N\right)\ni U^\dag=e^{-i x_aT^a}\rightarrow e^{-i z_aT^a}=e^{-i \left(x_a+i y_a\right)T^a}=U^{-1}\in\mathrm{SL}\left(N,\mathbb{C}\right).
$$
In the previous section we found the critical points of 
the action. We now see how to attach thimbles to them: 
this is the point at which the theory at hand
will reveal itself as simple. Not having a local gauge 
symmetry in place, the construction will go on quite smoothly:
we simply need to construct the union of the SA paths 
originating from critical points.

\noindent 
Having defined the Lie derivative (\ref{eq:nable})
we can write the SA equations as 
\beq \label{eq:SAgauge}
\frac{\mrm{d}}{\mrm{d}\tau}U_\muh\l(n;\tau\r)=\l(i\,T^a\bar{\nabla}_{n,\muh}^a\overline{S\l[U\l(\tau\r)\r]}\r)U_\muh\l(n;\tau\r).
\eeq
It is easy to show that the solutions of these equations display the main properties we
expect: the real part of the action is nondecreasing, while the
imaginary part stays constant. Namely, since $\frac{\mrm{d}}{\mrm{d}\tau}=
\bar{\nabla}_{n,\muh}^a\bar{S}\,\nabla_{n,\muh}^a+
\nabla_{n,\muh}^aS\,\bar{\nabla}_{n,\muh}^a$ we have that 
$$
\frac{\mrm{d}S^R}{\mrm{d}\tau}=\frac{1}{2}\frac{\mrm{d}}{\mrm{d}\tau}\l(S+\bar{S}\r)=\frac{1}{2}\l(\bar{\nabla}_{n,\muh}^a\bar{S}\,\nabla_{n,\muh}^a S+\nabla_{n,\muh}^aS\,\bar{\nabla}_{n,\muh}^a\bar{S}\r)=\left\Vert\nabla S\right\Vert^2\geq0
$$
and
$$
\frac{\mrm{d}S^I}{\mrm{d}\tau}=\frac{1}{2i}\frac{\mrm{d}}{\mrm{d}\tau}\l(S-\bar{S}\r)=\frac{1}{2i}\l(\bar{\nabla}_{n,\muh}^a\bar{S}\,\nabla_{n,\muh}^a S-\nabla_{n,\muh}^aS\,\bar{\nabla}_{n,\muh}^a\bar{S}\r)=0.
$$

\noindent
Starting from a critical point, we can now reach any point on the
thimble by integrating one particular SA equation. 
This is not the end of the story. In order to construct the tangent
space at each point of the thimble, we need to select a basis for the
tangent space at the critical point and then transport it along
the flow. Lie derivatives obey non-trivial commutation relations
\beqa
\nonumber\l[\nabla_{n,\muh}^a\,,\nabla_{m,\nuh}^b\r] &=& -f^{abc}\,\nabla_{n,\muh}^c\,\delta_{n,m}\delta_{\muh,\nuh}\\
\nonumber\l[\bar{\nabla}_{n,\muh}^a\,,\bar{\nabla}_{m,\nuh}^b\r] &=& -f^{abc}\,\bar{\nabla}_{n,\muh}^c\,\delta_{n,m}\delta_{\muh,\nuh}\\
\nonumber\l[\nabla_{n,\muh}^a\,,\bar{\nabla}_{m,\nuh}^b\r] &=& 0
\eeqa
from which we can get commutation relations for vectors 
$V\equiv
V_{n,\muh,a}\nabla_{n,\muh}^a+\bar{V}_{n,\muh,a}\bar{\nabla}_{n,\muh}^a$
$$
\l[V,V'\r]_{n,\muh,c}=-f^{abc}\,V_{n,\muh,a}V'_{n,\muh,b}.
$$
Taking $V'_{n,\muh,c}=\bar{\nabla}_{n,\muh}^c\bar{S}$ we can derive
the equation for transporting a vector $V$ from the
critical point to any point along the flow described by (\ref{eq:SAgauge})
\beq
\frac{\mrm{d}}{\mrm{d}\tau}V_{n,\muh,c}=\bar{\nabla}_{m,\nuh}^a\bar{\nabla}_{n,\muh}^c\bar{S}\,\bar{V}_{m,\nuh,a}+f^{abc}\,\bar{\nabla}_{n,\muh}^b\bar{S}\,V_{n,\muh,a}.
\eeq

\subsection{Takagi vectors at critical points} 
\label{sec:takagi}
Takagi's factorization provides the characterization of the thimble in
the vicinity of the critical point $p_\sigma$ (with coordinates
$z_\sigma$). We introduce the vector notation and expand the action to 
second order around $z_\sigma$

\begin{equation}
Z=
\begin{pmatrix}
z^1\\
\vdots\\
z^n
\end{pmatrix}
\in\mbb{C}^n
\qquad\qquad \qquad 
S(z)\approx S(z_\sigma)+\frac{1}{2}Z^TH(S;p_\sigma)Z\label{eq:action_expansion}
\end{equation}

\noindent  
where we have assumed $z_\sigma=0$ for the sake of
simplicity. Takagi's factorization theorem states that, given the
\emph{complex symmetric} matrix $H(S;p_\sigma)$ (the Hessian, in our case), there
exists a \emph{unitary} $n\times n$ matrix $W$ such that
$W^TH(S;p_\sigma)W=\Lambda$, with
$\Lambda=\mrm{diag}\l(\lambda_1,\cdots,\lambda_n\r)$ and the
$\lambda_i$ (called \emph{Takagi values}) are all real and
non-negative. We will find that in the case at hand they are all
positive. The columns of $W$ are $n$ normalized \emph{Takagi vectors} 
$v^{(i)}$, that is

\begin{equation*}
\sum_{k=1}^nv^{(i)}_k\bar{v}^{(j)}_k=\delta^{ij}
\end{equation*}

\noindent  
so that we can rephrase Takagi's theorem as

\begin{equation*}
H(S;p_\sigma)v^{(i)}=\lambda_i\bar{v}^{(i)}
\end{equation*}

\noindent  
or, equivalently,

\begin{equation*}
H(S;p_\sigma)W=\overline{W}\Lambda.
\end{equation*}

Takagi's vectors provides a basis for the tangent space at the
critical point. This also mean that each SA leaves the critical point
along a direction which is a given linear combination of 
Takagi's vectors. As we will see, our preferred way of singling out one
particular point on the thimble is indeed simple: we choose such a direction and then
specify when to stop while integrating the SA.

We saw we have three critical points: $\{U_k=e^{2\pi i k/3}\id\}$
($k=0,1,2$). After defining

\begin{equation*}
B_k\equiv 2\l[\cosh\l(\frac{\mu_c}{T}\r)+\cosh\l(\frac{\mu}{T}+\frac{2\pi i k}{3}\r)\r]
\end{equation*}

\noindent  
we have $S(U_k)=-3N_f\log B_k$ and $\nabla^b\nabla^aS(U)\bigr|_{U_k}=\lambda_ke^{i\,\varphi_k}\delta^{ab}$, with

\begin{equation*}
\lambda_ke^{i\,\varphi_k}\equiv N_f\l[B_k^{-1}\l(\cosh\l(\frac{\mu}{T}+\frac{2\pi i k}{3}\r)-2B_k^{-1}\sinh^2\l(\frac{\mu}{T}+\frac{2\pi i k}{3}\r)\r)\r]
\end{equation*}

We thus have one single Takagi value $\lambda_k$, while the $8$ Takagi
vectors are thus recognized as $v^{[k](i)}_j=e^{-i\,\varphi_k/2}\delta_{ij}$.

\subsection{Reflection symmetry and its consequences} 
\label{sec:symmYuya}

We have already stated that a contribution from each critical
point is expected: this is a direct consequence of the fact that 
they are all sitting on the original domain of integration\footnote{For
a discussion of which thimbles do contribute to the decomposition in 
(\ref{eq:basicO}) see \eg \cite{AUthimble}}. Collecting contributions 
of more than one thimble to solve the theory is in general a delicate
issue. We now see that symmetries can play a major role, providing
useful checks for results and even making our life simpler. These
observations were pointed out in \cite{yuya_symmetry}. 

We will now show that the action of 0+1 QCD fulfills a reflection symmetry: $\overline{S(A)}=S(-\bar{A})$ with $U=e^{iA}$. This ensures the \emph{reality} of the partition function (and of the expectation value of the Polyakov loop as well). This symmetry of the theory is manifestly fulfilled by the decomposition in thimbles and holds at every order in perturbation theory as well, so we shall recover it in the semiclassical expansion. Consider the QCD partition function

\begin{equation*}
Z_{N_f}(\mu)=\int\DD\psi\DD\bar{\psi}\DD U\,e^{-N_f\bar{\psi}D(U,\mu)\psi}=\int\DD U\,{\det}^{N_f}\l(D(U,\mu)\r)
\end{equation*}

The action (in our case, the Dirac determinant is the only component) is invariant under charge conjugation $\mcal{C}$ defined by 

\begin{align*}
\mcal{C}
\begin{cases}
\psi\rightarrow C^{-1}\bar{\psi}^T\\
\bar{\psi}\rightarrow-\psi^T C\\
U_\nuh(n)\rightarrow\bar{U}_\nuh(n) & \l(A_\nuh(n)\rightarrow-A^T_\nuh(n)=-\bar{A}\r)\\
\mu\rightarrow-\mu
\end{cases}
\end{align*}

with the matrix $C$ satisfying $C\gamma_\mu C^{-1}=-\gamma_\mu^T$. Thus, we can employ charge conjugation to substitute $\det D(U,\mu)\rightarrow\det D(\bar{U},-\mu)$ leaving the action invariant. We also recall the generalization of $\gamma_5$-hermiticity at finite chemical potential 

\begin{equation*}
\det D(U,-\mu)=\overline{\det D(U,\mu)}
\end{equation*}

This implies that

\begin{equation*}
\overline{S(A)}\sim\overline{\det D(U,\mu)}\overset{\gamma_5\text{-herm.}}{=}\det D(U,-\mu)\overset{\mcal{C}\text{-inv.}}{=}\det D(\bar{U},\mu)\sim S(-\bar{A})
\end{equation*}

We have shown that the aforementioned reflection symmetry is fulfilled
and thus we expect thimbles to appear in conjugate pairs. This is
indeed the case: consider the three critical points
$\{U_k\}$. $U_0=\id$ is real and therefore self-conjugate; the
consequence of this is that computations on the associated thimble
yield real results. As for the other two critical points, being
$e^{4\pi i/3}=e^{-2\pi i/3}$, we immediately see that
$U_2=\overline{U_1}$. This implies that $U_1$ and $U_2$ form a
conjugate pair of critical points and results of integration on $U_2$
should be the complex conjugate of those on $U_1$, yielding an overall
real contribution to the partition function (and also to the
expectation value of observables). The
chiral condensate and the quark number density automatically respect
this symmetry, being derivatives of the partition function. Notice that
the symmetry holds for the Polyakov loop (it is obvious, since
$\overline{\tr\,U}=\tr\,\bar{U}$) and anti-Polyakov loop as well. 
All this is well evident in numerical simulations: in figure 1 we show
an example in the case of the chiral condensate. In all the numerical
results that we present in the following we take advantage of this 
symmetry: results from thimble 1 and 2 are averaged. 
\begin{figure}[t]
\begin{center}
\includegraphics[height=11cm,clip=true]{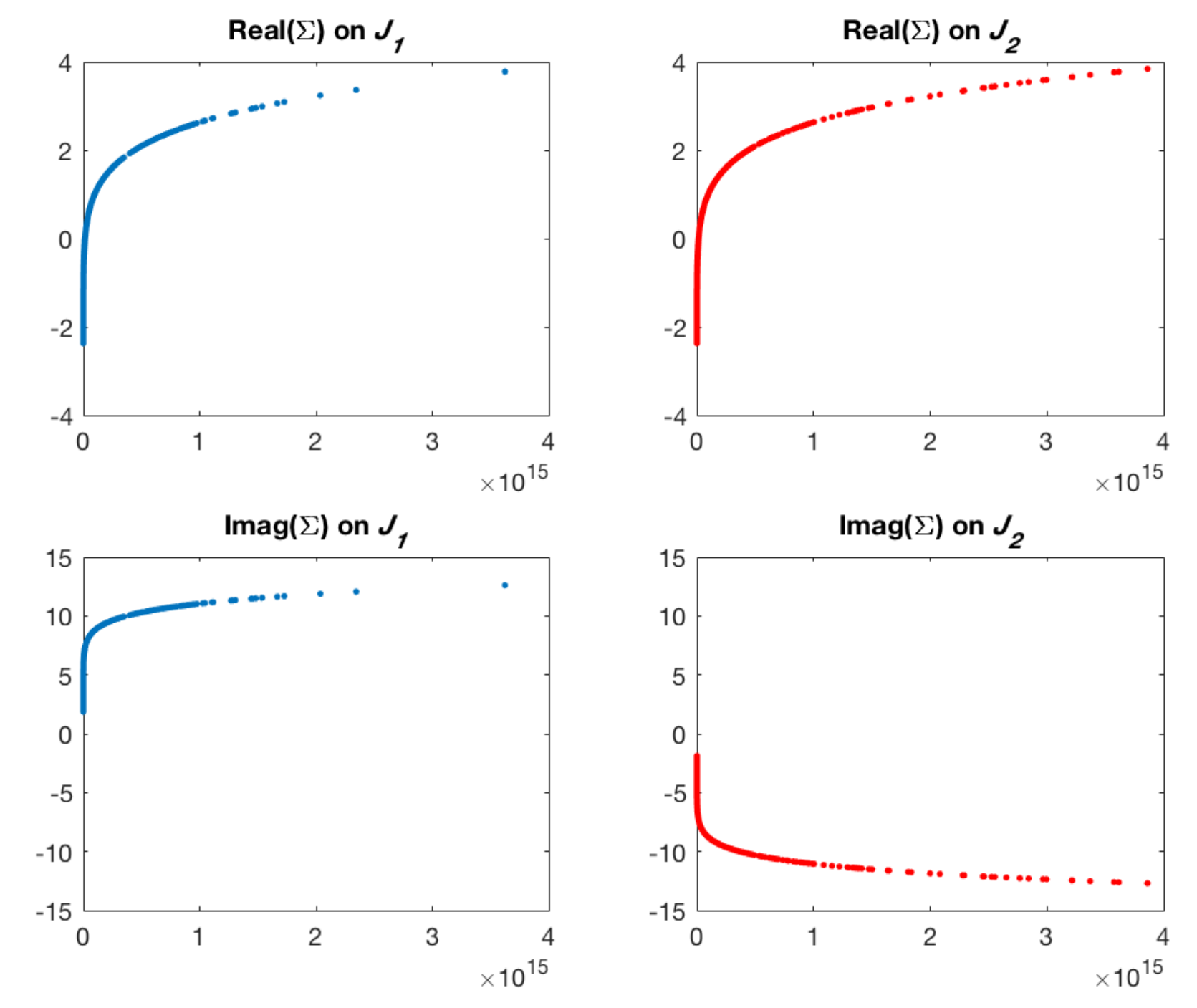}
\caption{Real and imaginary part of the chiral condensate on the
  thimbles attached to $U_1$ and $U_2$. The symmetry described in the
  text is evident. Results were obtained at $m=1, \, T=0.5, \, \mu=1.5, \,
  N_f=1$ and are plotted vs the values of $Z_\sigma$ (see eq (\ref{eq:totalZ_Zn}) later).}
\label{fig:SymmAtWork}
\end{center}
\end{figure}

\subsection{Semiclassical expansion}
\label{sec:semiclassics}

In this section we will compute semiclassical expansions around
thimbles in $0+1$ QCD. In general if in (\ref{eq:action_expansion})
we change variables according to $Z=W\eta$, the action in the $\eta$ variables becomes
\begin{equation}
S(\eta)=S(z_\sigma)+\frac{1}{2}\sum_{i=1}^n\lambda_i\eta_i^2+\cdots\label{eq:gaussS_eta}
\end{equation}

After setting $Z=W\eta$, so that $\dd^nz=\det W\dd^n\eta=e^{i\,\omega_\sigma}\dd^n\eta$, the partition function becomes
\begin{equation}
Z\approx\sum_{\sigma\in\Sigma}n_\sigma\,e^{-S(z_\sigma)}\,e^{i\,\omega_\sigma}\int\limits_{\mbb{R}^n}\dd^n\eta\,e^{-\frac{1}{2}\sum\limits_{i=1}^n\lambda_i\eta_i^2}=(2\pi)^{\frac{n}{2}}\sum_{\sigma\in\Sigma}n_\sigma\,\frac{e^{-S(z_\sigma)}}{\sqrt{\det\Lambda_\sigma}}\,e^{i\,\omega_\sigma}\label{eq:semicl_Z}
\end{equation}

In our case, we have $\det\Lambda_k=\lambda_k^8$ and
$e^{i\,\omega_k}=\l(e^{-i\,\varphi_k/2}\r)^8$, so that
(\ref{eq:semicl_Z}) reads

\begin{equation*}
Z\approx(2\pi)^4\sum_{k=0,1,2}n_k\,e^{3N_f\log B_k}\lambda_k^{-4}\,e^{-4i\,\varphi_k}
\end{equation*}

As for the expectation value of an observable $O$, we can expand $O(z)$ around $p_\sigma$
\begin{equation*}
O(z)\approx O(z_\sigma)+{\nabla_Z}^T OZ+\frac{1}{2}Z^T H_\sigma^O Z
\end{equation*}
with
\begin{equation*}
{\nabla_Z}^T O\equiv\l(\frac{\de O}{\de z_1}\biggr|_{z_\sigma},\cdots,\frac{\de O}{\de z_n}\biggr|_{z_\sigma}\r)
\end{equation*}
and
\begin{equation*}
\l(H_\sigma^O\r)_{ij}\equiv\frac{\de^2 O}{\de z_i\de z_j}\biggr|_{z_\sigma}
\end{equation*}
It is obvious that, in general, $[H(S;p_\sigma),H_\sigma^O]\neq 0$ and therefore we cannot expect $W$ to ``diagonalize'' both $H(S;p_\sigma)$ and $H_\sigma^O$. The expectation value of the observable is given by
\begin{equation*}
\langle O\rangle\approx\frac{1}{Z}\sum_{\sigma\in\Sigma}n_\sigma\,e^{-S(z_\sigma)}\,e^{i\,\omega_\sigma}\int\limits_{\mbb{R}^n}\dd^n\eta\l(O(z_\sigma)+{\nabla_Z}^TOW\eta+\frac{1}{2}\eta^TC_\sigma^O\eta\r)e^{-\frac{1}{2}\sum\limits_{i=1}^n\lambda_i\eta_i^2}
\end{equation*}
with $C_\sigma^O\equiv W^TH_\sigma^OW$. The first term in the expansion of $O$ comes out of the integral, giving the partition function itself (actually, the contribution of the thimble $\mcal{J}_\sigma$). The second term is linear in $\eta_i$ and therefore gives no contribution to the Gaussian integral. For the same reason, the third term contributes only with terms which are quadratic in $\eta_i$, that is when
\begin{equation*}
\eta^TC_\sigma^O\eta=\sum_{i=1}^n\sum_{j=1}^n\l(C_\sigma^O\r)_{ij}\eta_i\eta_j\rightarrow\sum_{i=1}^n\l(C_\sigma^O\r)_{ii}\eta_i^2
\end{equation*}
Therefore we need to compute only the diagonal terms of $C_\sigma^O$ and, after performing a Gaussian integral
we arrive at
\begin{equation}
\langle O\rangle\approx\frac{1}{Z}(2\pi)^{\frac{n}{2}}\sum_{\sigma\in\Sigma}n_\sigma\,\frac{e^{-S(z_\sigma)}}{\sqrt{\det\Lambda_\sigma}}\,e^{i\,\omega_\sigma}\l(O(z_\sigma)+\frac{1}{2}\sum\limits_{i=1}^n\frac{\l(C_\sigma^O\r)_{ii}}{\lambda_i}\r)\label{eq:semicl_O}
\end{equation}

\noindent
The expectation value of the Polyakov loop can be computed starting from expression (\ref{eq:semicl_O})

\begin{equation*}
\langle\tr\,U\rangle\approx\frac{1}{Z}(2\pi)^4\sum_{k=0,1,2}n_k\,e^{3N_f\log B_k}\lambda_k^{-4}\,e^{-4i\,\varphi_k}\l(\tr\,U_k+\frac{1}{2}\frac{1}{\lambda_k}\sum_{i=1}^8\l(C_k^{\tr\,U}\r)_{ii}\r)
\end{equation*}

where
\begin{align*}
&\l(C_k^{\tr\,U}\r)_{ii}=\sum_{j=1}^8\sum_{l=1}^8\l(H_k^{\tr\,U}\r)_{jl}v^{(i)}_jv^{(i)}_l=e^{-i\,\varphi_k}\l(H_k^{\tr\,U}\r)_{ii}=e^{-i\,\varphi_k}\nabla^i\nabla^i\tr\,U\bigr|_{U_k}\\
&=-e^{-i\,\varphi_k}e^{2\pi ik/3}\tr\l(T^iT^i\,\id\r)=-\frac{1}{2}e^{-i\,\varphi_k}e^{2\pi ik/3}
\end{align*}

Being $\tr\,U_k=e^{2\pi ik/3}\,\tr\,\id=3\,e^{2\pi ik/3}$, it follows that
\begin{equation*}
\tr\,U_k+\frac{1}{2}\frac{1}{\lambda_k}\sum_{i=1}^8\l(C_k^{\tr\,U}\r)_{ii}=e^{2\pi ik/3}\l(3-\frac{2}{\lambda_k}e^{-i\,\varphi_k}\r)
\end{equation*}

and finally
\begin{equation*}
\langle\tr\,U\rangle\approx\frac{1}{Z}(2\pi)^4\sum_{k=0,1,2}n_k\,e^{3N_f\log B_k}\lambda_k^{-4}\,e^{-4i\,\varphi_k+2\pi ik/3}\l(3-\frac{2}{\lambda_k}e^{-i\,\varphi_k}\r)
\end{equation*}

From the previous considerations on the reflection symmetry featured by $0+1$ QCD, we can see that reality of $Z$ and $\langle\tr\,U\rangle$ is achieved by setting $n_1=n_2$. This is so since the contribution of $\mcal{J}_2$ to $Z$ and $\langle\tr\,U\rangle$ is the complex conjugate of the contribution of $\mcal{J}_1$. This is manifest in the semiclassical expansion thanks to $S(U_2)=\overline{S(U_1)}$, $B_2=\bar{B}_1$, $\lambda_2=\lambda_1$, $e^{i\,\varphi_2}=\overline{e^{i\,\varphi_1}}=e^{-i\,\varphi_1}$, all following from $e^{4\pi i/3}=e^{-2\pi i/3}=\overline{e^{2\pi i/3}}$. Thus we can rephrase $Z$ as
\begin{equation*}
Z\approx Z_0+Z_1+Z_2
\end{equation*}

with $Z_0\in\mbb{R}$ and $Z_2=\bar{Z}_1$ (so that $\l|Z_1\r|=\l|Z_2\r|$). The semiclassical expansion on thimbles also provides an easy way to compute an estimate for the relevance of $\mcal{J}_{1,2}$ with respect to $\mcal{J}_0$ in the computation of \eg$\,$the partition function. We define the relative weight $r^{1,2}_0$
\begin{equation}
r^{1,2}_0\equiv\frac{\l|Z_{1,2}\r|}{\l|Z_0\r|}=\frac{\l|e^{3N_f\log B_{1,2}}\r|\lambda_{1,2}^{-4}}{\l|e^{3N_f\log B_0}\r|\lambda_0^{-4}}=\l(\frac{\lambda_{1,2}}{\lambda_0}\r)^{-4}\l|\frac{B_{1,2}}{B_0}\r|^{3N_f}
\label{eq:qcd01_r120}
\end{equation}

and study it at different values of $\frac{\mu}{T}$ and $m$. This, as we shall see, provides a reliable estimate which can be compared with the results of numerical simulations. We note that, being $B_0=A+2\cosh(\mu/T)\in\mbb{R}$ and $B_1=A-\cosh(\mu/T)+i\,\sqrt{3}\sinh(\mu/T)$
\begin{align*}
&\l|B_{1,2}\r|^2=A^2+\cosh^2\l(\frac{\mu}{T}\r)-2A\cosh\l(\frac{\mu}{T}\r)+3\sinh^2\l(\frac{\mu}{T}\r)\\
&=A^2+4\cosh^2\l(\frac{\mu}{T}\r)-2A\cosh\l(\frac{\mu}{T}\r)-3<A^2+4\cosh^2\l(\frac{\mu}{T}\r)+4A\cosh\l(\frac{\mu}{T}\r)=\l|B_0\r|^2
\end{align*}

so that
\begin{equation*}
r^{1,2}_0\underset{N_f\rightarrow\infty}{\longrightarrow}0
\end{equation*}

for any value of $\frac{\mu}{T}$ and $m$ (the ratio
$\lambda_{1,2}/\lambda_0$ is independent on $N_f$). One thus expects
that integrating only over $\mcal{J}_0$ will give more accurate
results at large $N_f$, \ie {\em there is a regime where the leading
thimble dominance scenario actually shows up}. This is of course 
a semiclassical estimate: the reliability of this prediction will 
be checked against numerical simulations.

\section{Monte Carlo computations on thimbles}

Our preferred way of characterising points on a thimble goes through 
a constructive approach, which we now recall in the formalism which is 
valid for any generic theory. Given a critical point, we 
saw in subsection \ref{sec:takagi} how to determine the tangent space. By 
performing the Takagi factorization of the Hessian we were left with 
Takagi values ${\lambda_i>0}$ and Takagi vectors
$v^{(i)}$, which provide a basis for the tangent space. The 
tangent space contains all the directions along which the SA
paths defined by\footnote{We denote by $t$ the {\em time} coordinate
parametrizing the flow along the SA path.}
\begin{equation}
\frac{d}{dt}z_i = \frac{\partial \bar{S}}{\partial \bar{z}^i}
\label{eq:SAeq}
\end{equation}
leave the critical point. 
If we impose a normalization condition
\[
\sum_{i=1}^nn_i^2=\mathcal{R}
\]
all those directions are mapped to vectors 
\[
\sum_{i=1}^nn_iv^{(i)}.
\]
It is thus
quite natural to single out any given point on a thimble by the
correspondence
\begin{equation}
\label{eq:nNt}
\mathcal{J}_\sigma\ni z\leftrightarrow \left(\hat{n},t\right)\in S^{n-1}_{\mathcal{R}}\times\mathbb{R}
\end{equation}
with $S^{n-1}_{\mathcal{R}}$ the $(n-1)$-sphere of radius
$\sqrt{\mathcal{R}}$. In \cite{thimbleCRM} we made use of this
approach to solve a Chiral Random Matrix Model by means of thimble
regularization. We now recall how to make use of (\ref{eq:nNt}) to
rephrase the integrals in the thimble decomposition of the
path integral. A change of variables in the integrals in
(\ref{eq:basicO}) will be involved. Let us first of all define
\begin{equation} 
Z_\sigma=\int\limits_{\mathcal{J}_\sigma}\mathrm{d}^ny\,e^{-S_R}\label{eq:single_thimble_Z}
\end{equation}
In the previous formula 
$\mathrm{d}^ny$ stands for the real volume form 
on the thimble $\mathcal{J}_\sigma$ (\ie the thimble attached to critical 
point $p_{\sigma}$). 
With a slight abuse of terminology we will refer to this expression 
as a partition function, which can be rewritten
\begin{equation}
Z_\sigma=\int\mathcal{D}\hat{n}\,Z_{\hat{n}}^{(\sigma)}\label{eq:totalZ_Zn}
\end{equation}
in terms of the measure over $S^{n-1}_{\mathcal{R}}$
\[
\mathcal{D}\hat{n}\equiv\prod_{k=1}^n\mathrm{d} n_k\delta\left(\left|\vec{n}\right|^2-\mathcal{R}\right)
\]
and the \emph{partial} partition functions
\begin{equation}
Z_{\hat{n}}^{(\sigma)}=\int\limits_{-\infty}^{+\infty}\mathrm{d} t\,\Delta_{\hat{n}}^{(\sigma)}(t)\,e^{-S_R(\hat{n},t)}\label{eq:partialZ}.
\end{equation}
The partition function $Z_\sigma$ has been decomposed in contributions
$Z_{\hat{n}}^{(\sigma)}$ attached to SA paths (\ie complete flow lines).
For each direction $\hat{n}$, a factor $\Delta_{\hat{n}}^{(\sigma)}(t)$ is left 
over after changing variables in the integral. It can be thought 
of as an extra contribution 
to the measure (on top of $e^{-S_R(\hat{n},t)}$) along the SA singled
out by the direction $\hat{n}$. 
The computation of $\Delta_{\hat{n}}^{(\sigma)}(t)$ is non-trivial: it is 
required that one parallel-transports the basis of
the tangent space at the critical point along the flow, to have a
basis $V_{\sigma}^{(i)}(\hat{n},t)$ at the (generic) point associated
to direction $\hat{n}$ and flow time $t$. More 
precisely, by assembling the $V_{\sigma}^{(i)}$ into the matrix $V_{\sigma}$, one finds
that 
\begin{equation} 
Z_{\hat{n}}^{(\sigma)}=2\sum_{i=1}^n\lambda_i^{(\sigma)}n_i^2\int\limits_{-\infty}^{+\infty}\mathrm{d} t\,e^{-S_{\mathrm{eff}}^{(\sigma)}(\hat{n},t)}\label{eq:partialZ_Seff}
\end{equation}
where the ${\lambda_i^{(\sigma)}>0}$ are the Takagi values (solutions
of the Takagi problem at the critical point $p_\sigma$) and 
the effective action $S_{\mathrm{eff}}^{(\sigma)}$ is given by
\begin{equation} 
S_{\mathrm{eff}}^{(\sigma)}(\hat{n},t)=S_R(\hat{n},t)-\log\left|\det V_{\sigma}(\hat{n},t)\right|\label{eq:Zn_Seff}.
\end{equation}
At the same time, the phase of $\det V_{\sigma}(\hat{n},t)$ provides
the residual phase $e^{i\omega(\hat{n},t)}$. 
We can now go back to (\ref{eq:basicO}) and rewrite it in terms of the
$\left(\hat{n},t\right)$ variables. Notice that at this point
$Z_\sigma$ will be irrelevant: it will reappear later on. All in
all we have
\begin{equation}
\label{eq:Integrals_nt}
\langle O \rangle = \frac{\sum_{\sigma} n_{\sigma} \,
  e^{-i\,S_I\left(p_{\sigma}\right)} \int_{\sigma}\mathcal{D}\hat{n}\; 
2\sum_{i=1}^n\lambda_i^{(\sigma)}n_i^2
\int\limits_{-\infty}^{+\infty}\mathrm{d}
t\,e^{-S_{\mathrm{eff}}^{(\sigma)}(\hat{n},t)} 
\,O (\hat{n},t)\, e^{i\omega (\hat{n},t)}}{\sum_{\sigma} n_{\sigma} \,
  e^{-i\,S_I\left(p_{\sigma}\right)} \int_{\sigma}\mathcal{D}\hat{n}\; 
2\sum_{i=1}^n\lambda_i^{(\sigma)}n_i^2
\int\limits_{-\infty}^{+\infty}\mathrm{d}
t\,e^{-S_{\mathrm{eff}}^{(\sigma)}(\hat{n},t)} \, e^{i\omega
  (\hat{n},t)}}
\end{equation}
Before proceeding we make a couple of observations:
\begin{itemize}
\item The expression in (\ref{eq:partialZ_Seff}) is not the same
  appearing in \cite{thimbleCRM}: the two are equivalent\footnote{One 
could say the way we proceed in \cite{thimbleCRM} makes the appearance
of $\Delta_{\hat{n}}(t)$ and the computation of the latter natural; the
expression for $\Delta_{\hat{n}}(t)$ we use here is easier to
deal with in practice.}.  
\item The notation $\int_{\sigma}$ could look generic: it reflects the
  fact that (\ref{eq:nNt}) holds {\em for each critical point}. On the
  other side, at each critical point one has to solve a different
  Takagi problem, resulting in different Takagi values
  $\lambda_i^{(\sigma)}$ and different Takagi vectors $v_\sigma^{(i)}$, which are the
  initial values for different $V_{\sigma}^{(i)}(\hat{n},t)$ (and this
  results in the end in 
  different $\Delta_{\hat{n}}^{(\sigma)}(t)$).
\end{itemize}

\subsection{Simulations by flat, crude Monte Carlo}

We can make use of flat, crude Monte Carlo to compute
the integrals in (\ref{eq:Integrals_nt}). The recipe is very simple
\begin{itemize}
\item We pick up randomly (with flat distribution) a direction $\hat{n}$. 
\item Since we want to compute the contribution coming from the SA leaving
  the critical point $p_\sigma$ along $\hat{n}$, we 
  prepare convenient {\em initial conditions} both for the field and
  for the tangent space basis vectors for such a SA. We can do this, 
since near the critical point
  solutions of (\ref{eq:SAeq}) are know as\footnote{For details see
    \eg \cite{thimbleCRM}.}
\begin{eqnarray}
z_j\left(t\right) & \approx & z_{\sigma,j}+\sum\limits_{i=1}^n
                              n_i  \; v_{\sigma j}^{\left(i\right)} \;
                              e^{\lambda_i^{(\sigma)} t} \nonumber \\
 V_{\sigma j}^{(i)}\left(t\right) & \approx &
v_{\sigma j}^{\left(i\right)} \; e^{\lambda_i^{(\sigma)} t} \nonumber
\end{eqnarray}
which we can compute for $t=t_0\ll0$.
\item We then integrate the SA equations for the field and 
the equations for transporting the basis vectors all the way up till 
we reconstruct the $\int\limits_{-\infty}^{+\infty}\mathrm{d}t$ 
integrals appearing in (\ref{eq:Integrals_nt})\footnote{Notice 
that while ascending we compute both the integral in the numerator and the
one in the denominator.}.
\end{itemize}

\begin{figure}[ht]
  \centering
  \subfloat[$m=0.1$, $N_f=1$]{
    \label{fig:qcd1_cc_m01_1}
    \includegraphics[height=2.5cm]{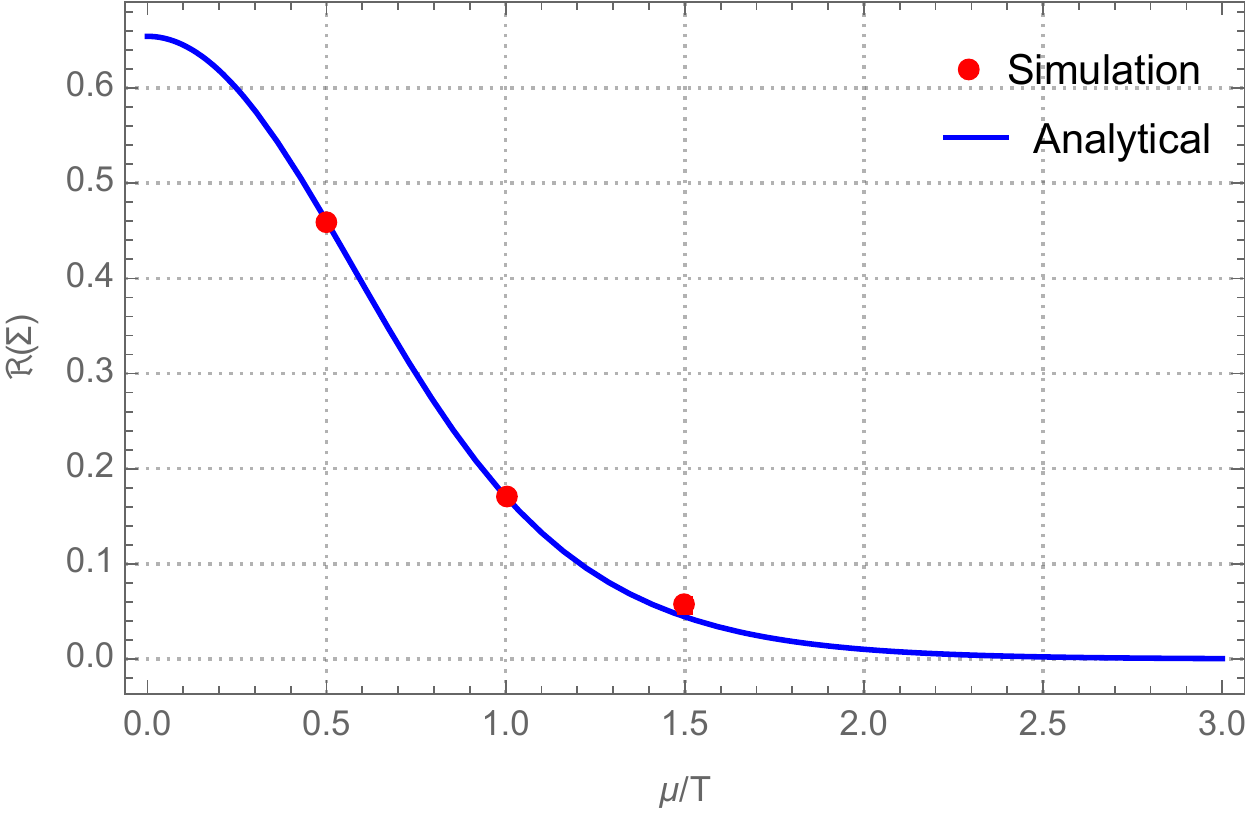}
  }
  \subfloat[$m=0.1$, $N_f=2$]{
    \label{fig:qcd1_cc_m01_2}
    \includegraphics[height=2.5cm]{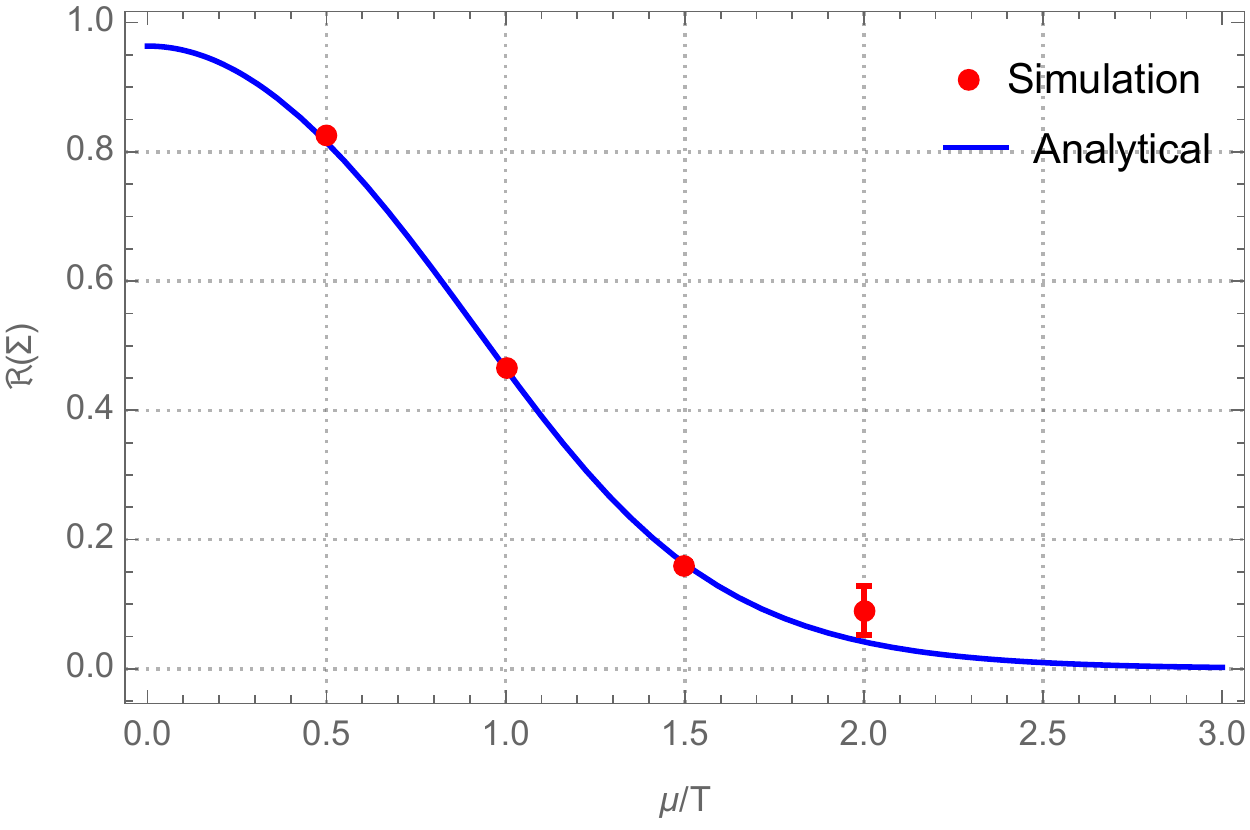}
  }
  \subfloat[$m=0.1$, $N_f=3$]{
    \label{fig:qcd1_cc_m01_3}
    \includegraphics[height=2.5cm]{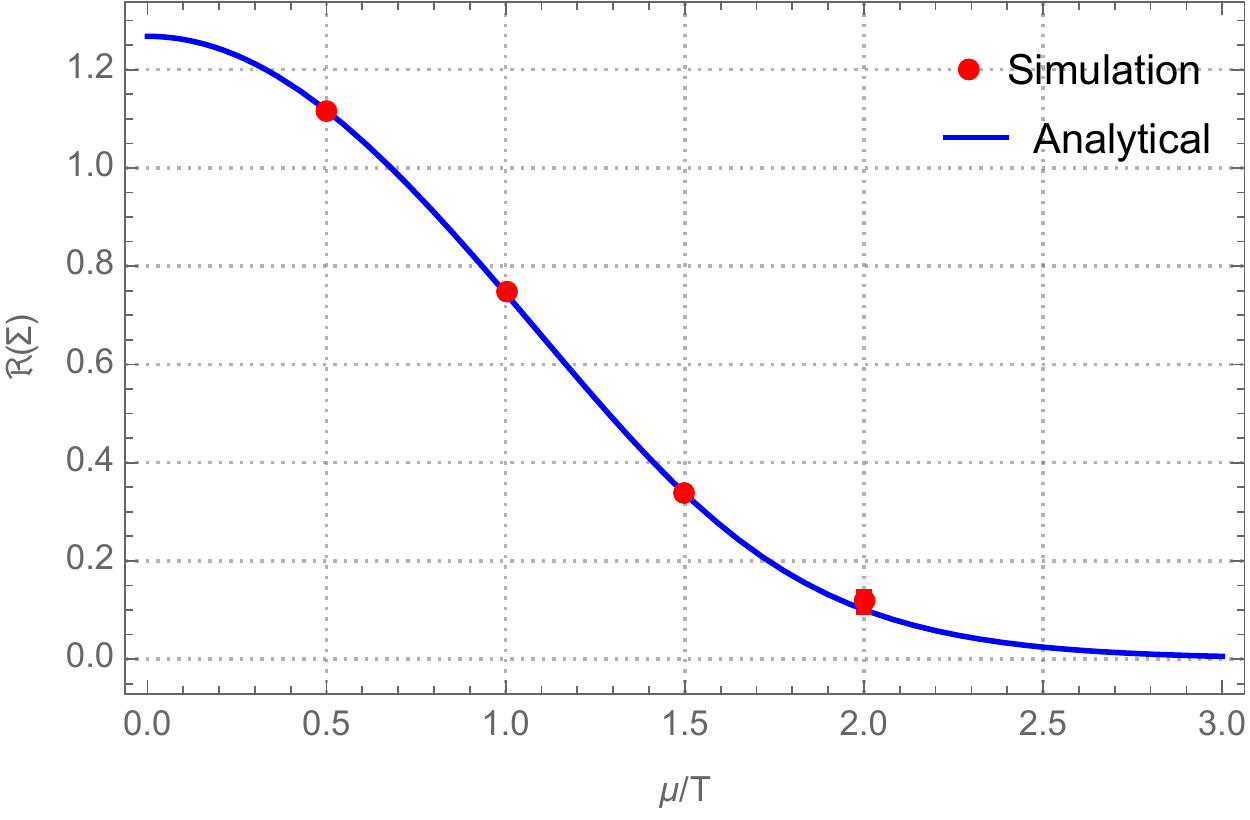}
  }
  \subfloat[$m=0.1$, $N_f=4$]{
    \label{fig:qcd1_cc_m01_4}
    \includegraphics[height=2.5cm]{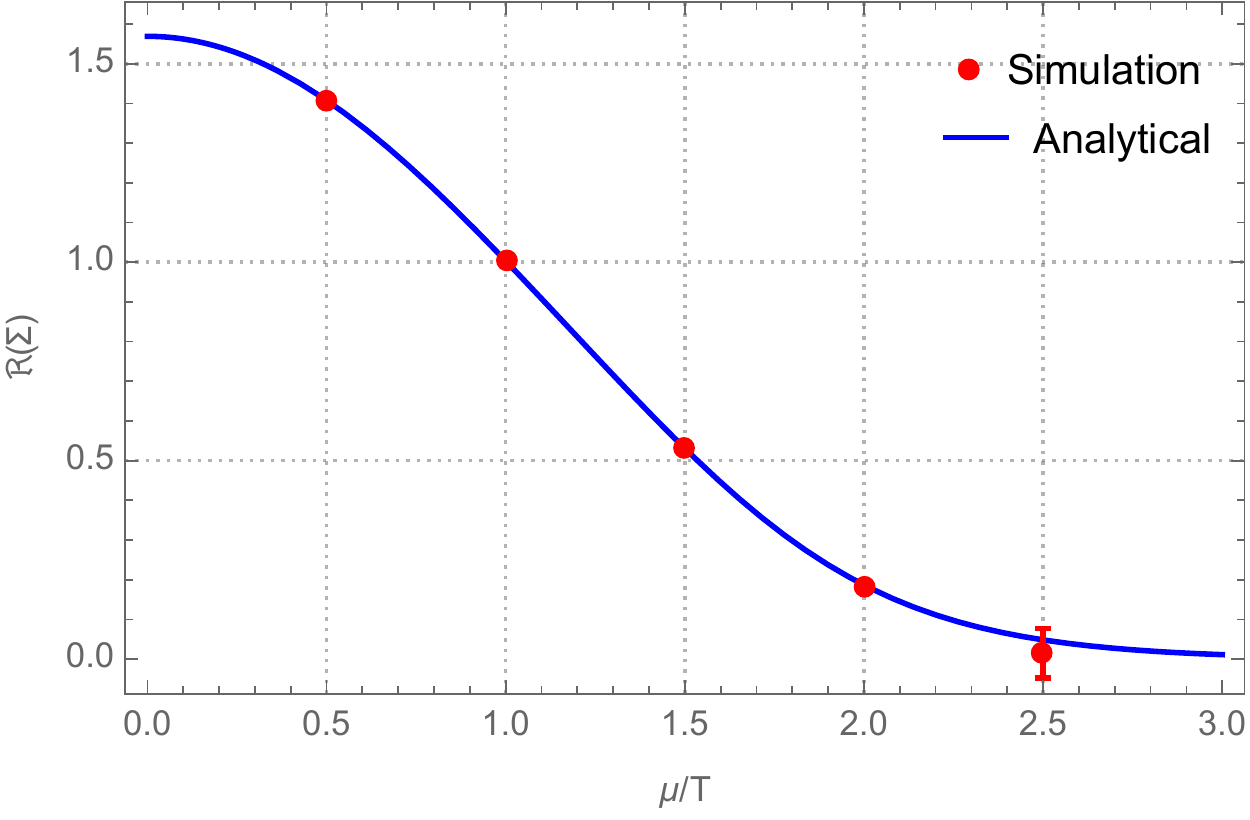}
  }
  \\
  \centering
  \subfloat[$m=0.1$, $N_f=1$]{
    \label{fig:qcd1_tru_m01_1}
    \includegraphics[height=2.5cm]{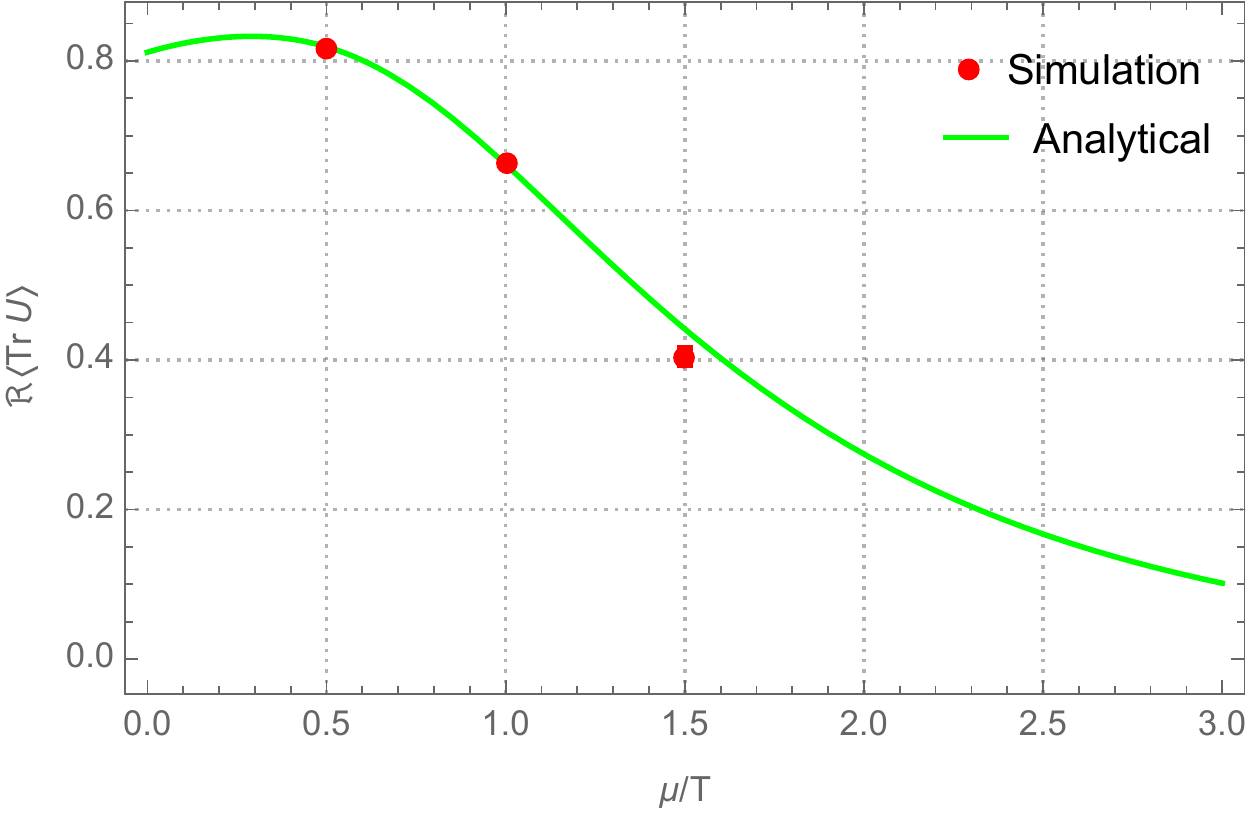}
  }
  \subfloat[$m=0.1$, $N_f=2$]{
    \label{fig:qcd1_tru_m01_2}
    \includegraphics[height=2.5cm]{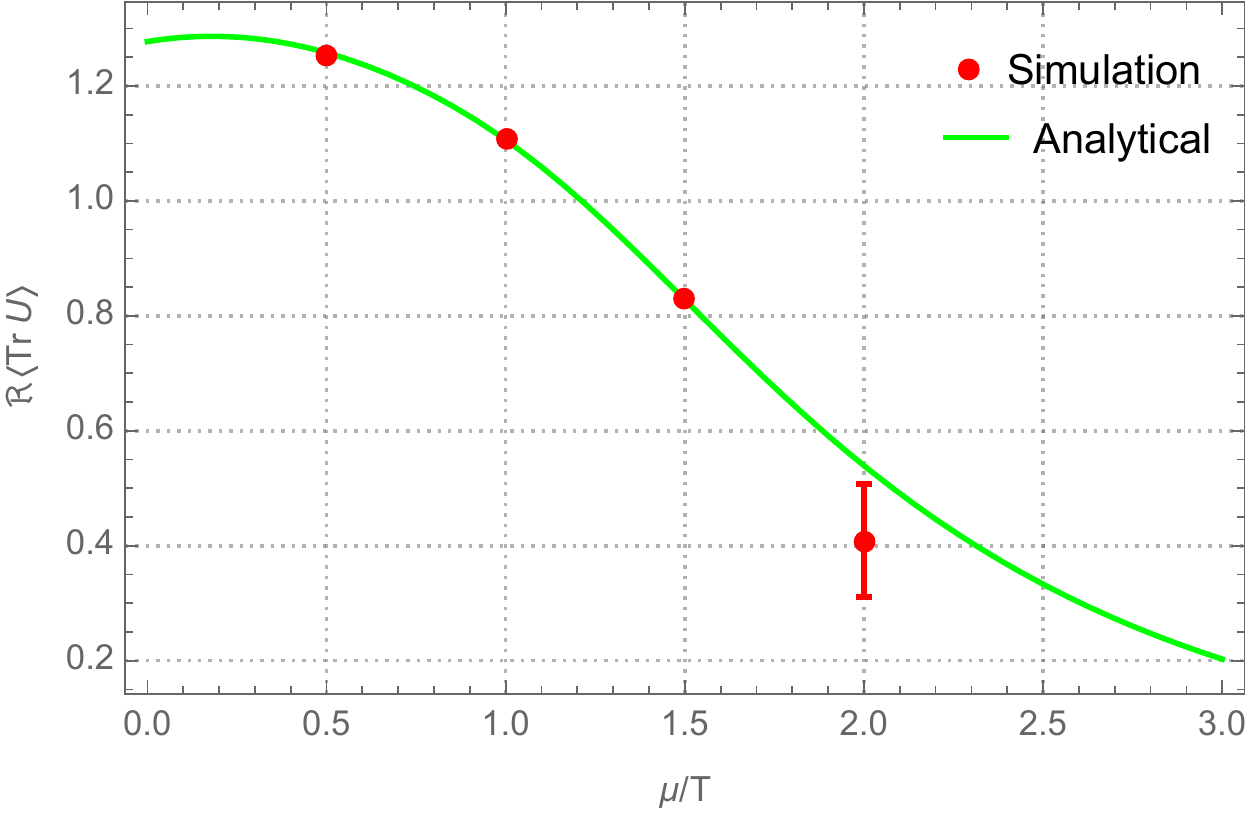}
  }
  \subfloat[$m=0.1$, $N_f=3$]{
    \label{fig:qcd1_tru_m01_3}
    \includegraphics[height=2.5cm]{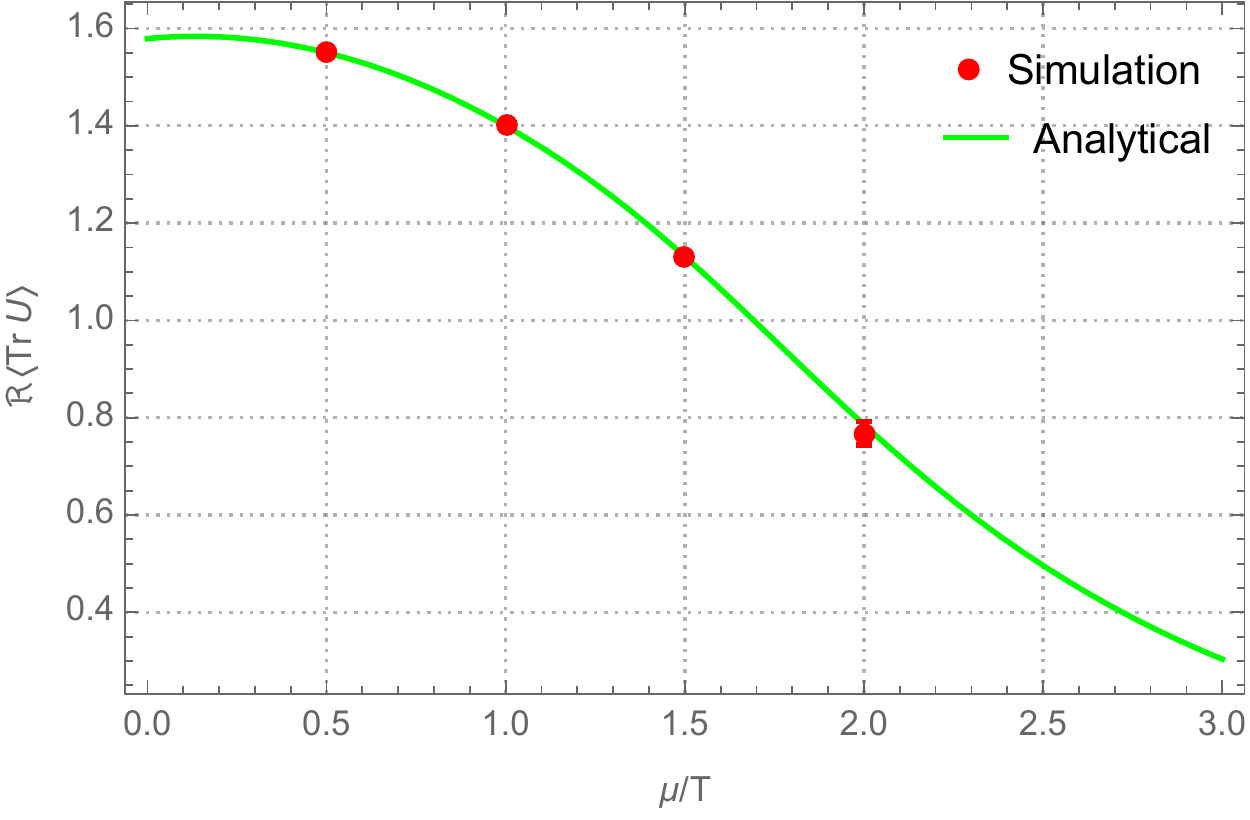}
  }
  \subfloat[$m=0.1$, $N_f=4$]{
    \label{fig:qcd1_tru_m01_4}
    \includegraphics[height=2.5cm]{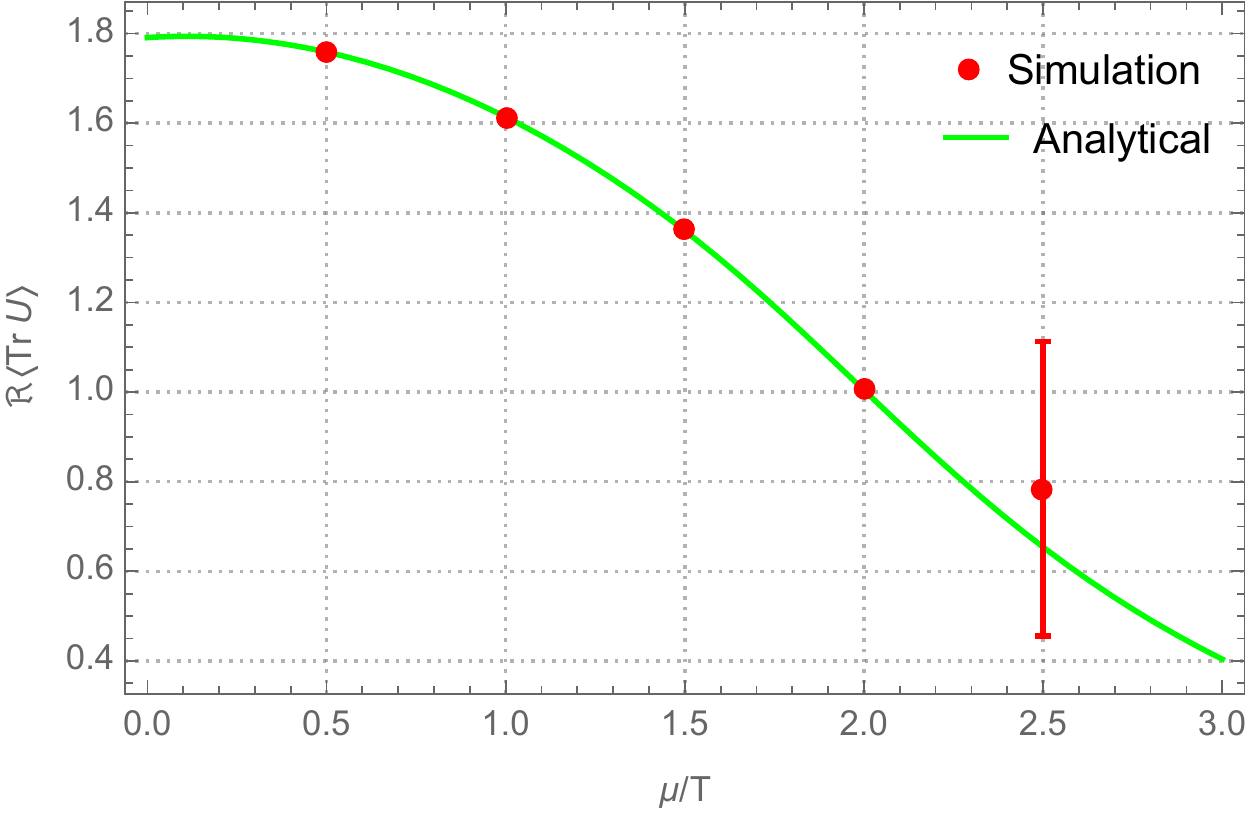}
  }
  \\
  \centering
  \subfloat[$m=1$, $N_f=1$]{
    \label{fig:qcd1_cc_m01_1}
    \includegraphics[height=2.5cm]{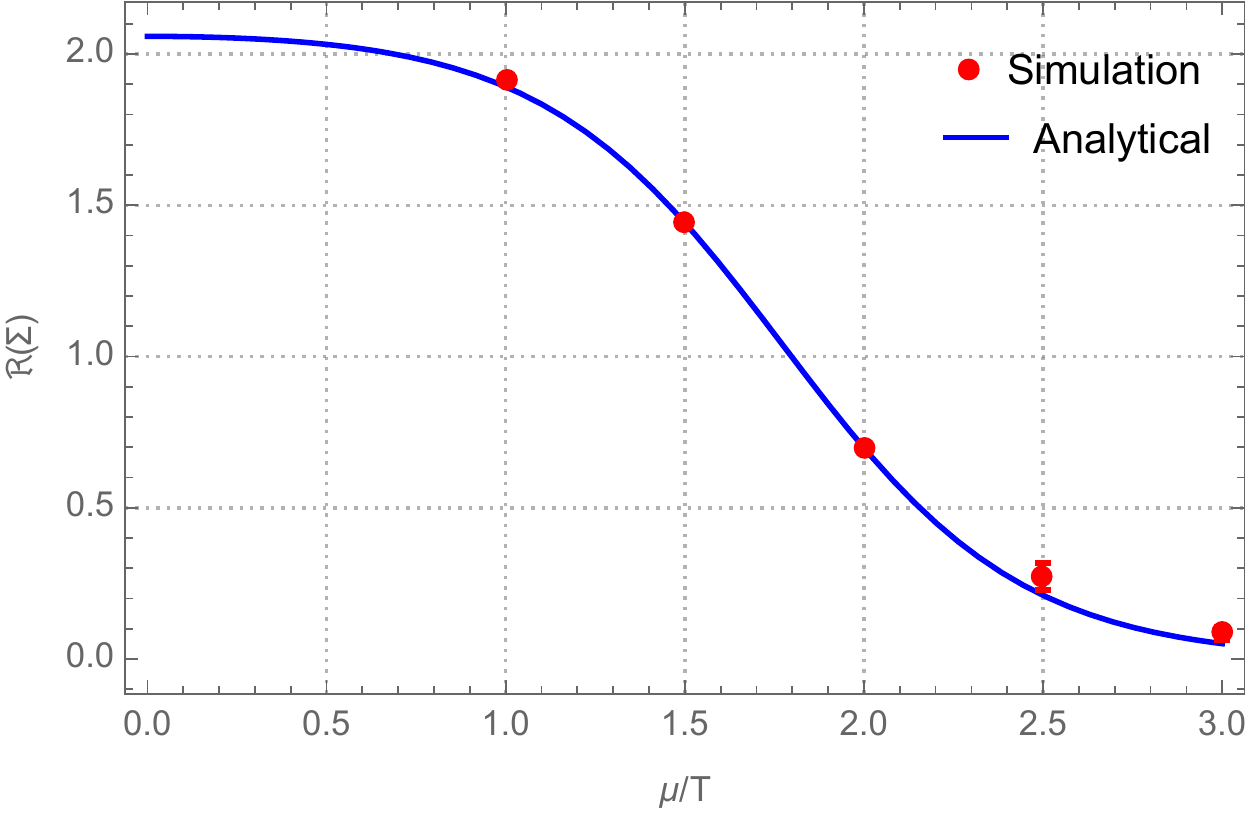}
  }
  \subfloat[$m=1$, $N_f=2$]{
    \label{fig:qcd1_cc_m01_2}
    \includegraphics[height=2.5cm]{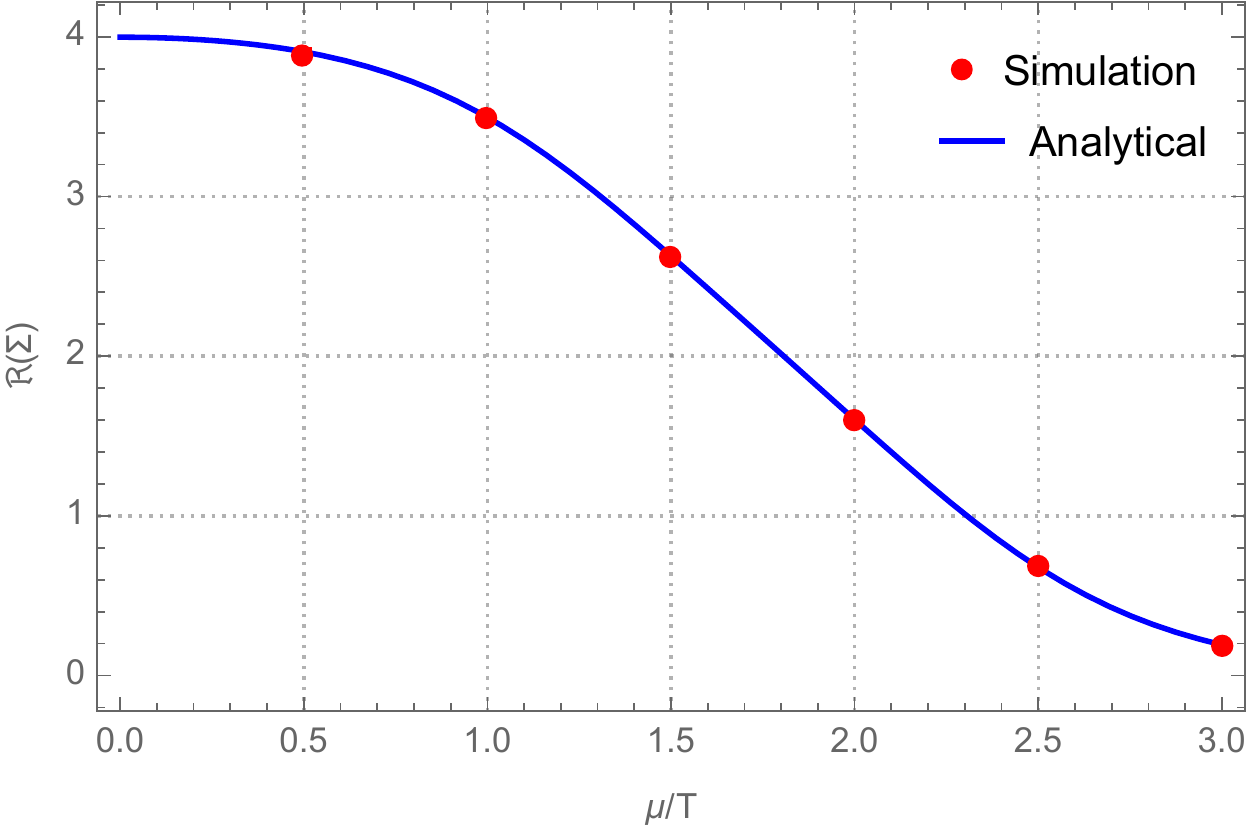}
  }
  \subfloat[$m=1$, $N_f=3$]{
    \label{fig:qcd1_cc_m01_3}
    \includegraphics[height=2.5cm]{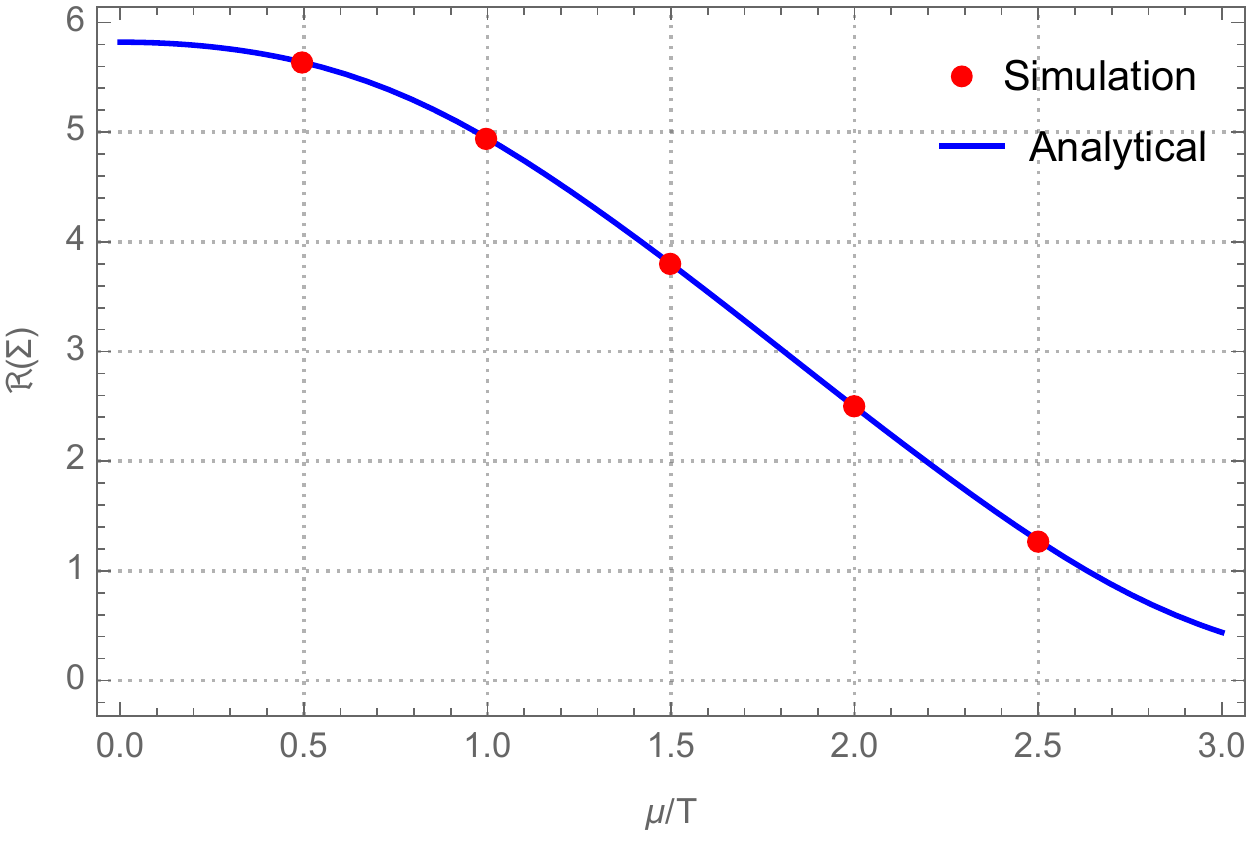}
  }
  \subfloat[$m=1$, $N_f=4$]{
    \label{fig:qcd1_cc_m01_4}
    \includegraphics[height=2.5cm]{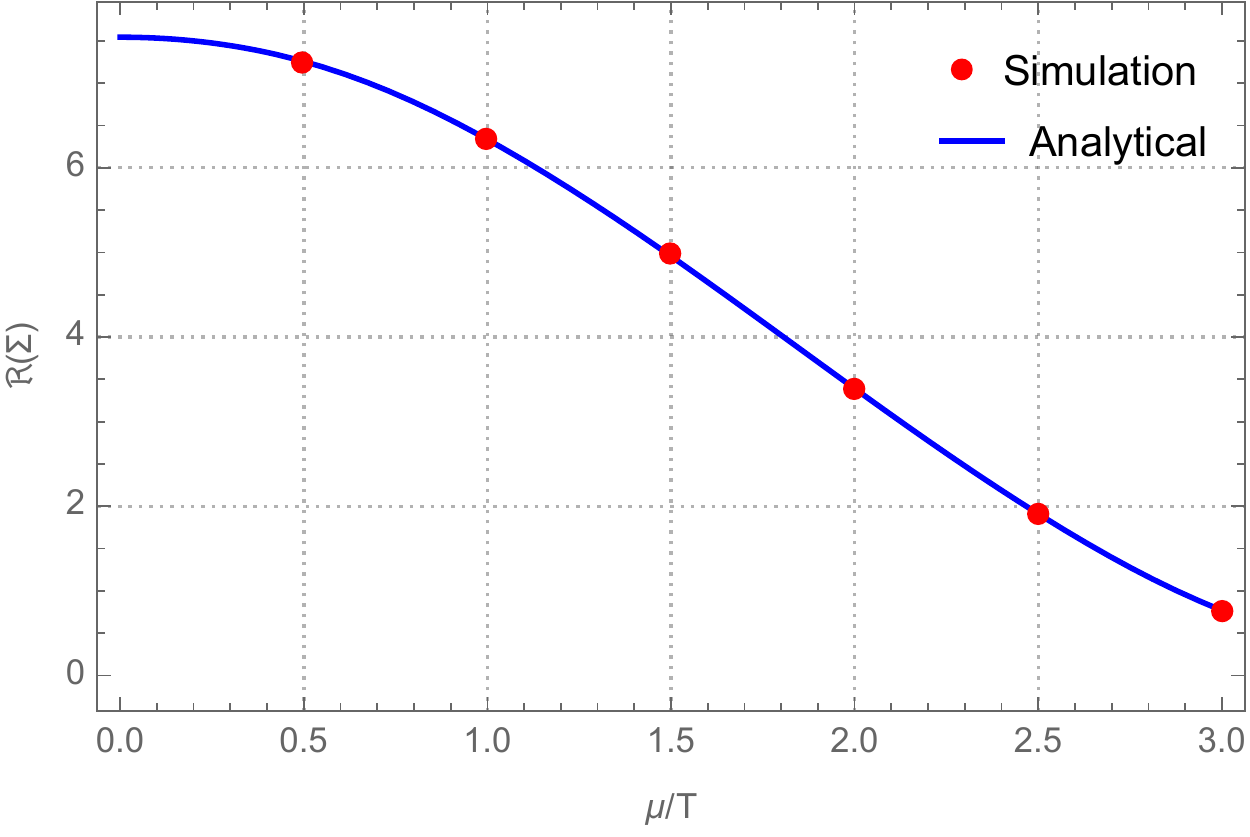}
  }
  \\
  \centering
  \subfloat[$m=1$, $N_f=1$]{
    \label{fig:qcd1_tru_m1_1}
    \includegraphics[height=2.5cm]{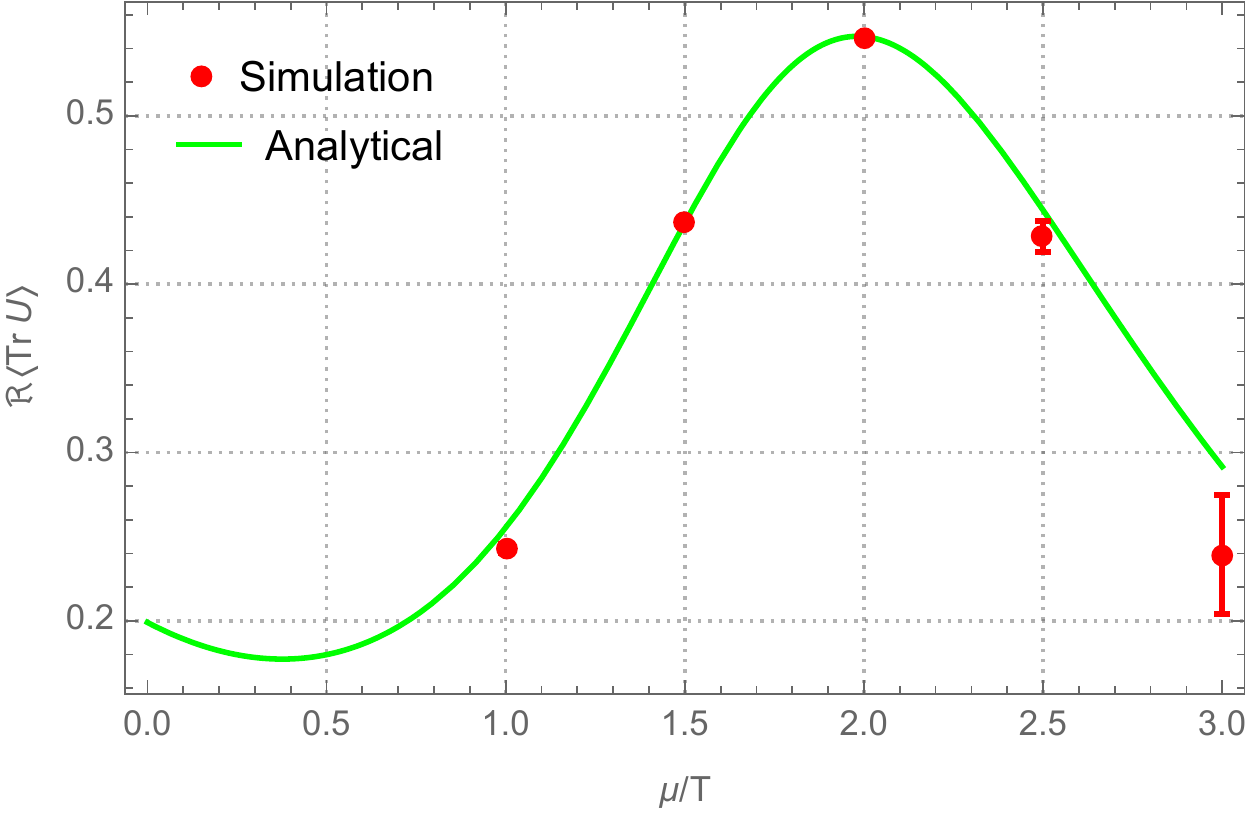}
  }
  \subfloat[$m=1$, $N_f=2$]{
    \label{fig:qcd1_tru_m1_2}
    \includegraphics[height=2.5cm]{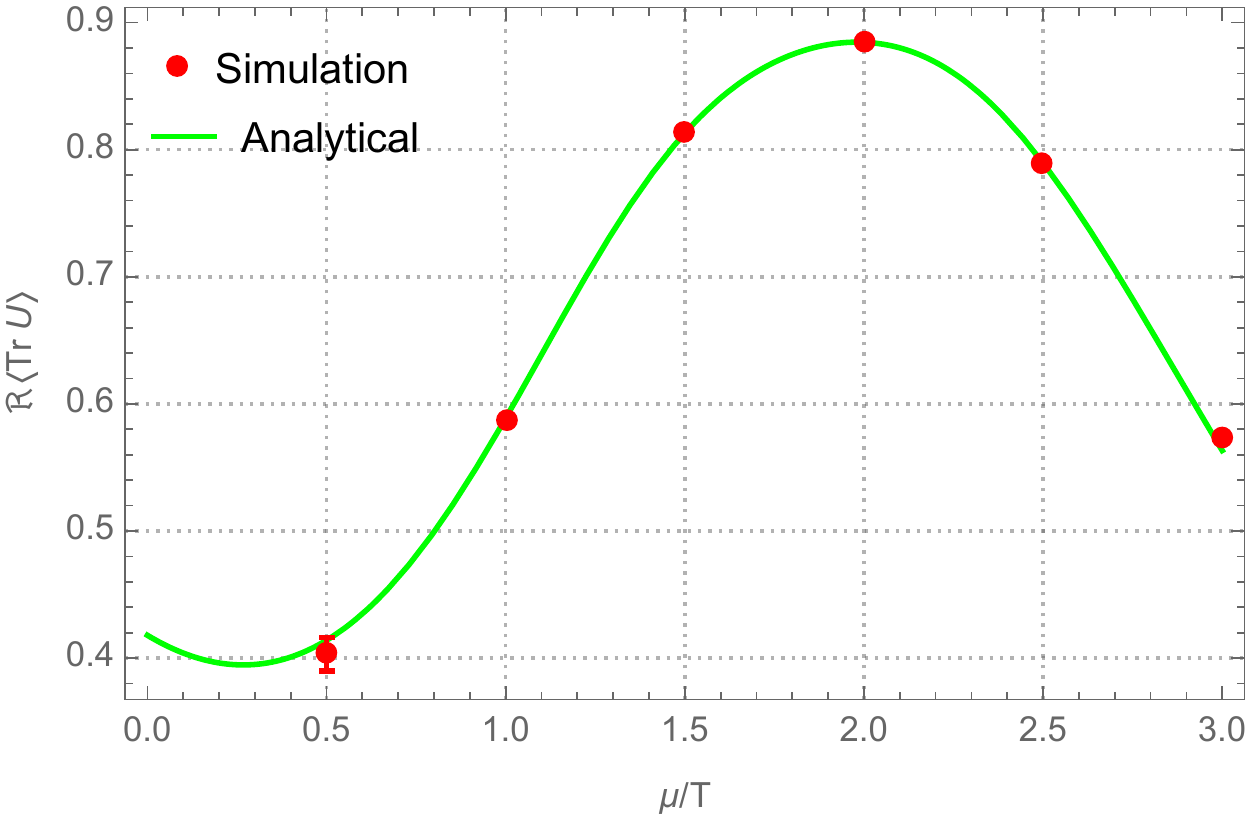}
  }
  \subfloat[$m=1$, $N_f=3$]{
    \label{fig:qcd1_tru_m1_3}
    \includegraphics[height=2.5cm]{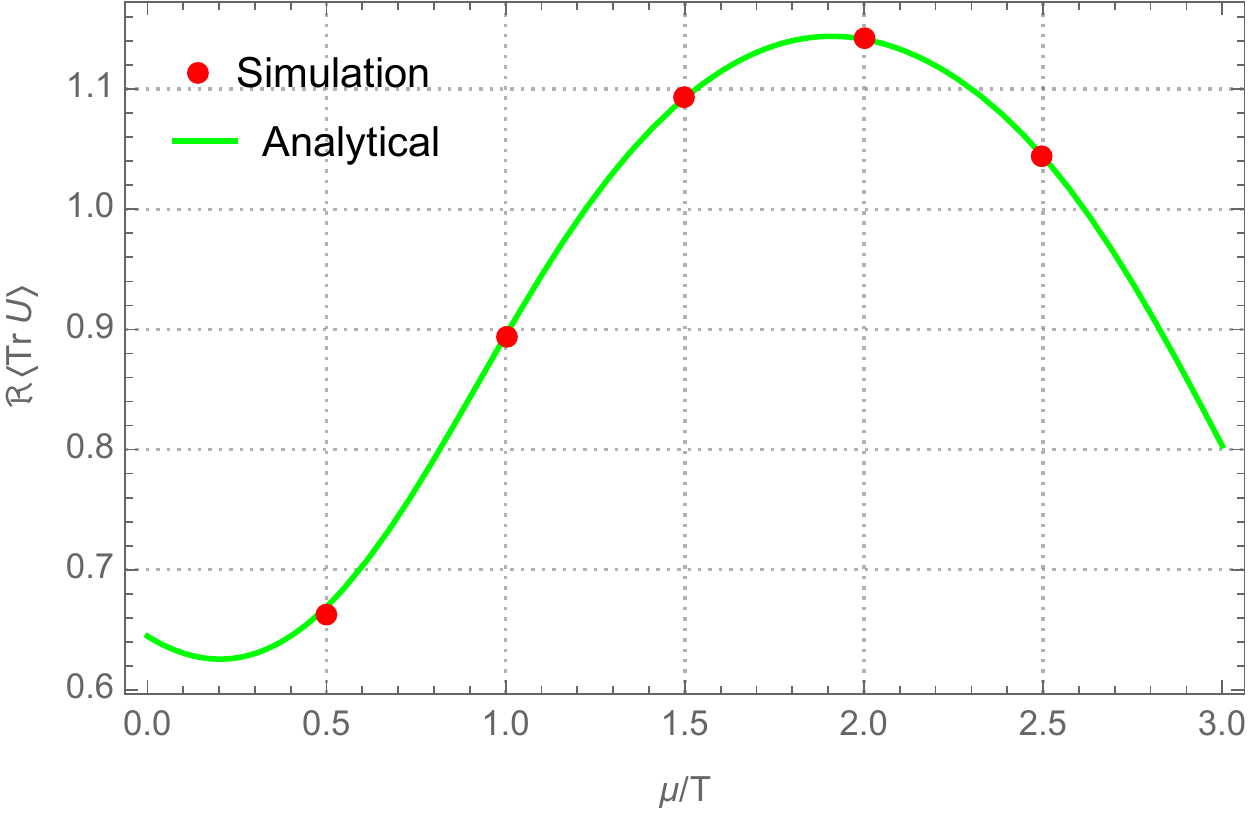}
  }
  \subfloat[$m=1$, $N_f=4$]{
    \label{fig:qcd1_tru_m1_4}
    \includegraphics[height=2.5cm]{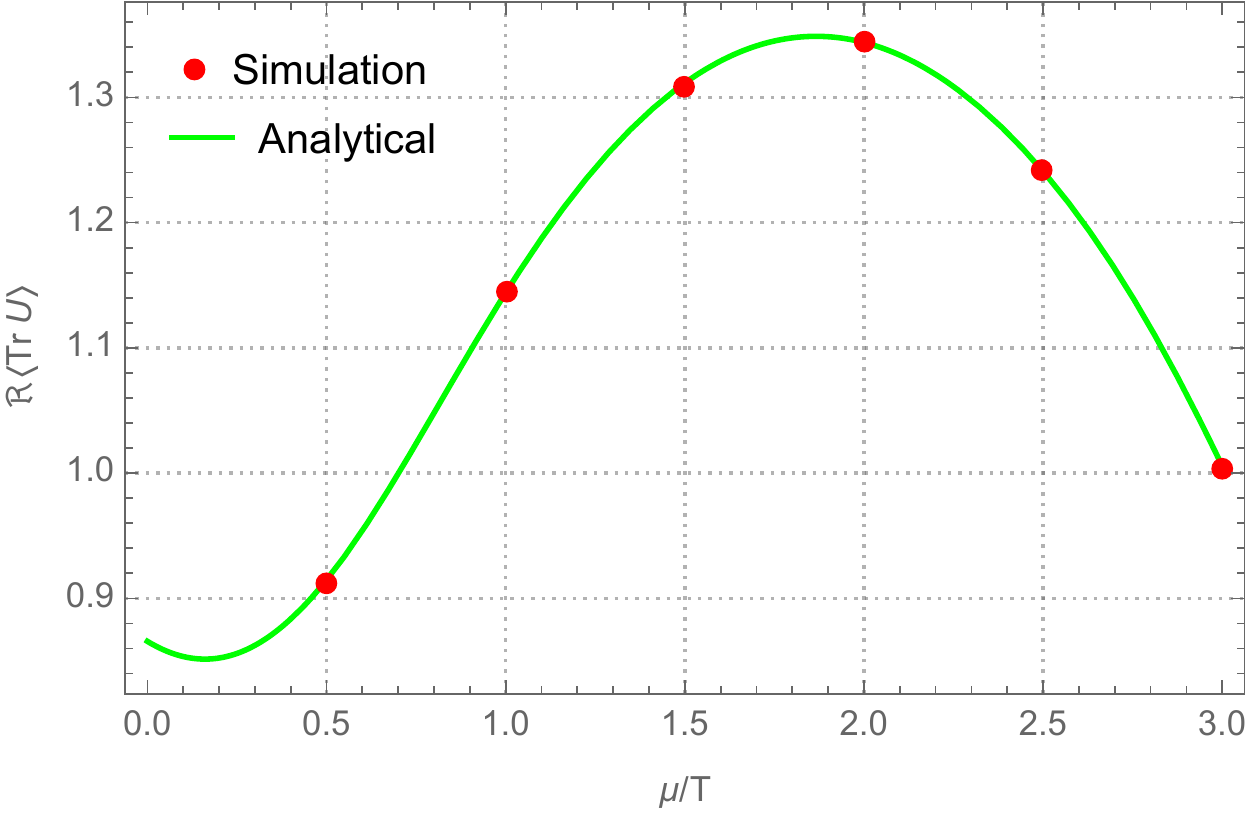}
  }
  \caption{Chiral condensate (blue; 1st and 3rd rows) and Polyakov loop
    (green; 2nd and 4th rows) expectation value for 0+1 QCD at $T=0.5$,
    $m=0.1$ (1st and 2nd rows) and $m=1$ (3rd and 4th rows) for
    $N_f=1,2,3,4$. Observables are plotted vs $\mu/T$.}
  \label{fig:qcd1_Nf1.2.3.4}
\end{figure}

\begin{figure}[ht]
  \centering
  \subfloat[$m=0.1$, $N_f=5$]{
    \label{fig:qcd1_cc_m01_5}
    \includegraphics[height=2.5cm]{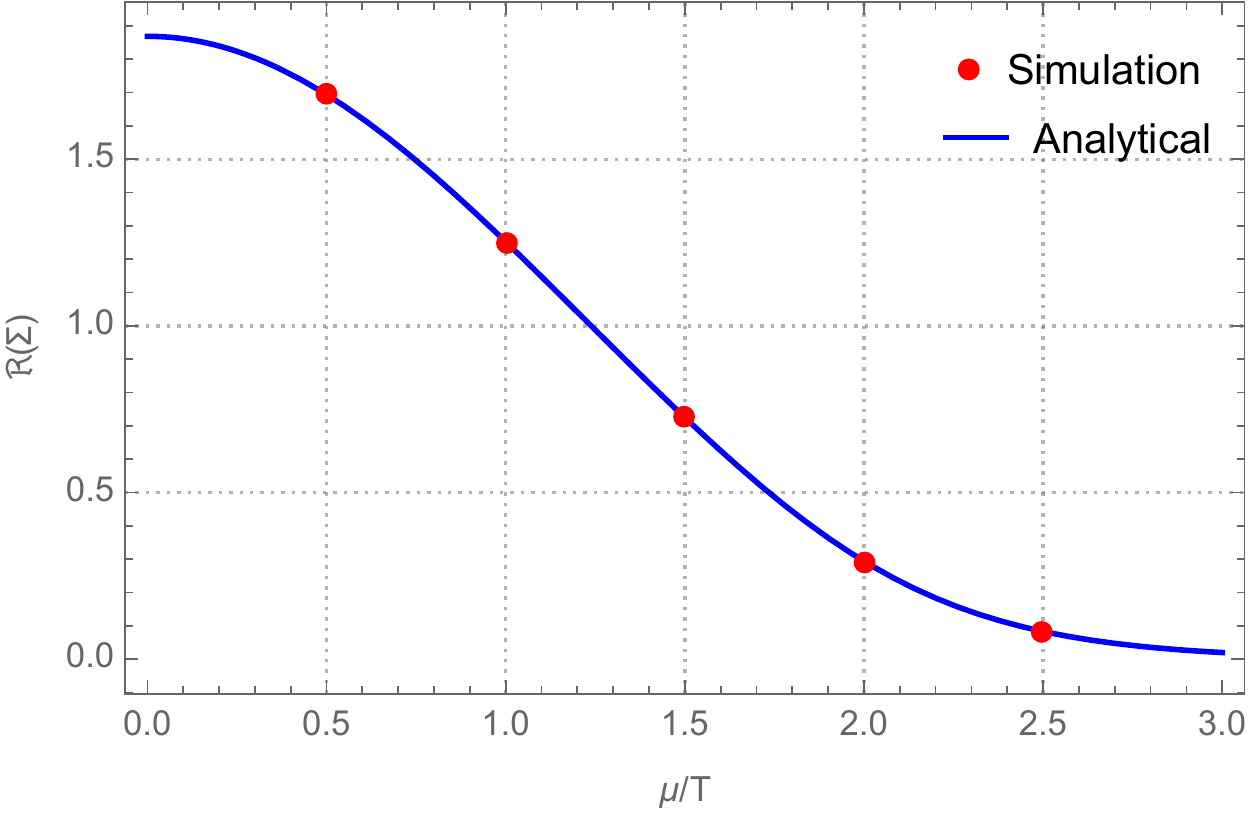}
  }
  \subfloat[$m=0.1$, $N_f=6$]{
    \label{fig:qcd1_cc_m01_6}
    \includegraphics[height=2.5cm]{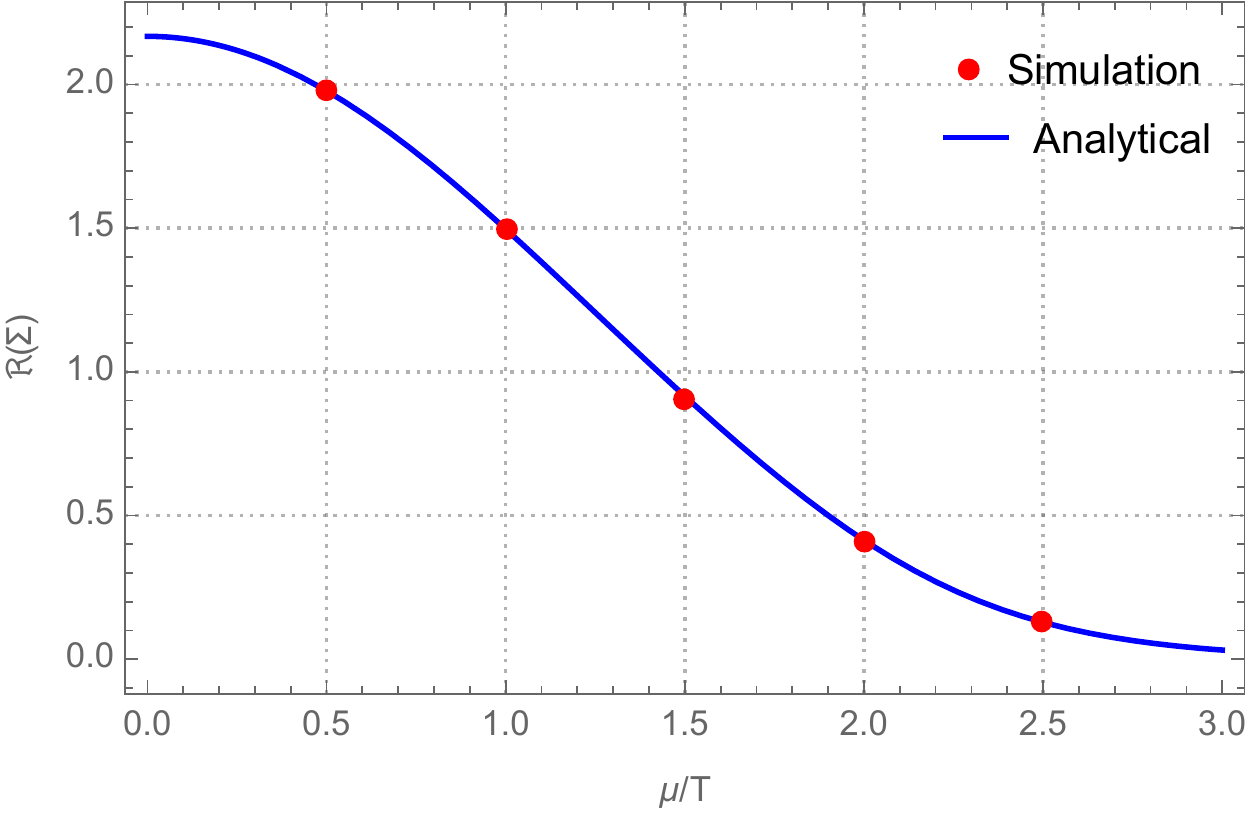}
  }
  \subfloat[$m=0.1$, $N_f=12$]{
    \label{fig:qcd1_cc_m01_12}
    \includegraphics[height=2.5cm]{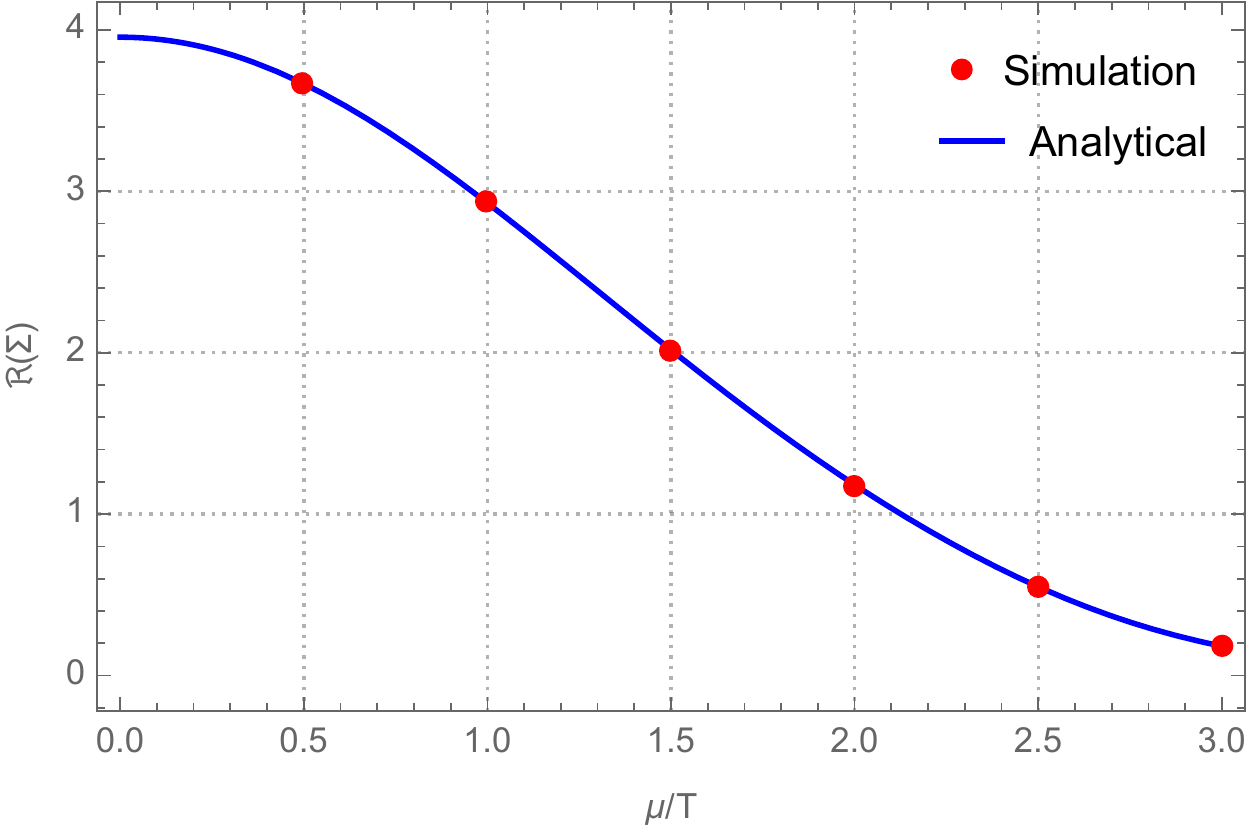}
  }
  \\
  \centering
  \subfloat[$m=0.1$, $N_f=5$]{
    \label{fig:qcd1_tru_m01_5}
    \includegraphics[height=2.5cm]{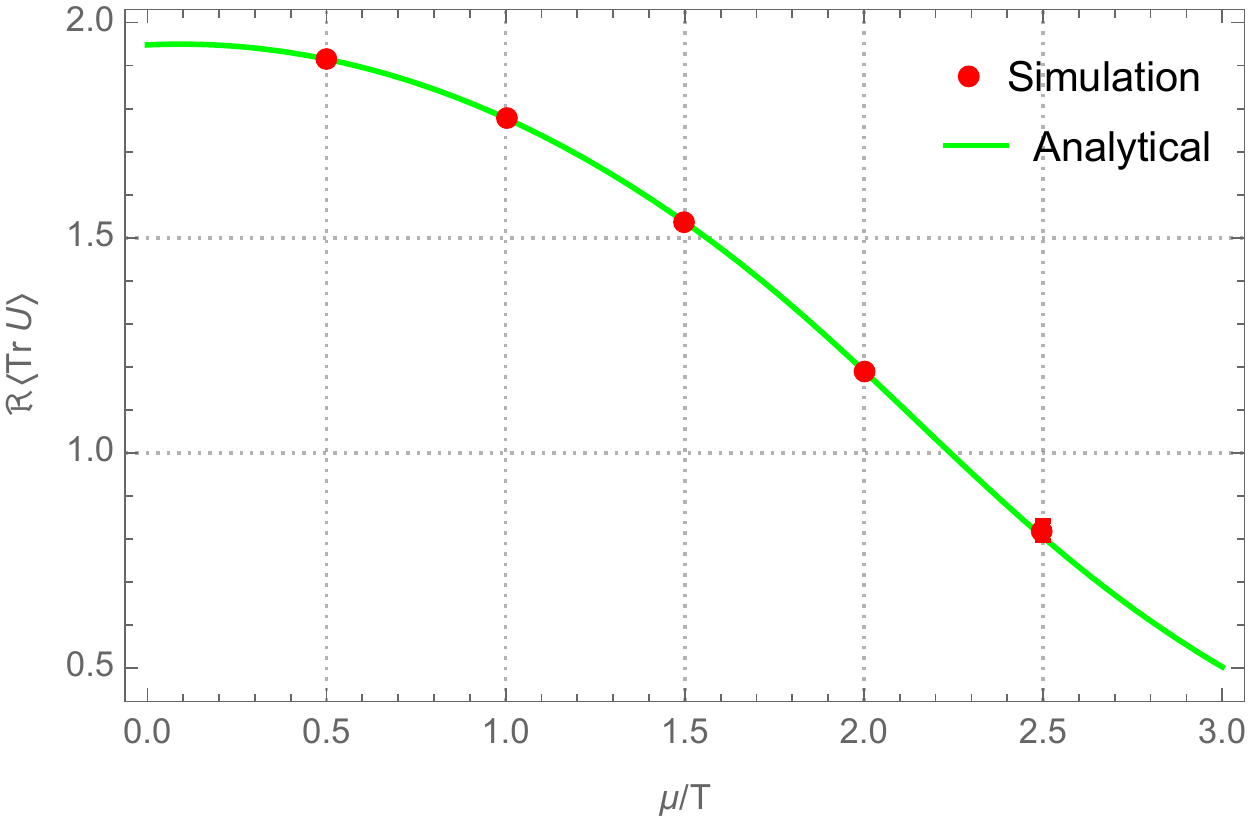}
  }
  \subfloat[$m=0.1$, $N_f=6$]{
    \label{fig:qcd1_tru_m01_6}
    \includegraphics[height=2.5cm]{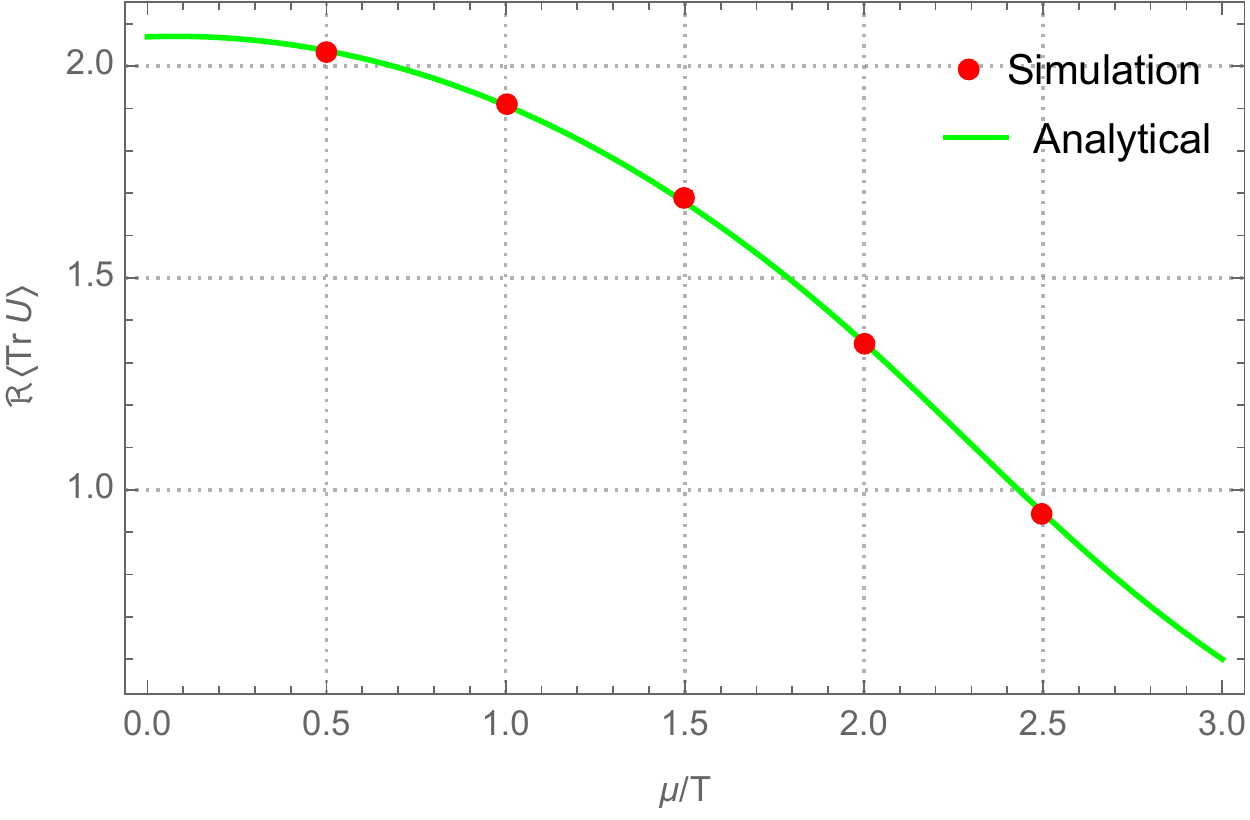}
  }
  \subfloat[$m=0.1$, $N_f=12$]{
    \label{fig:qcd1_tru_m01_12}
    \includegraphics[height=2.5cm]{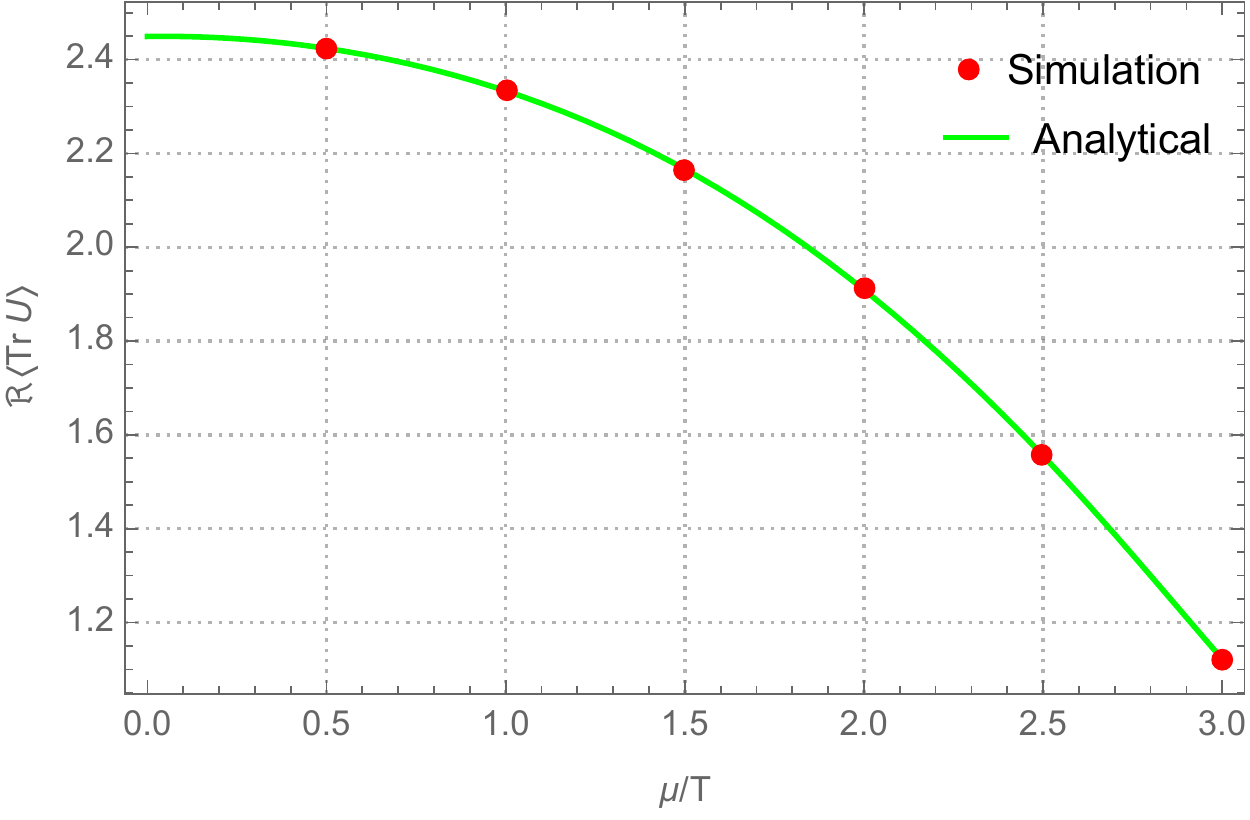}
  }
  \\
  \centering
  \subfloat[$m=1$, $N_f=5$]{
    \label{fig:qcd1_cc_m01_5}
    \includegraphics[height=2.5cm]{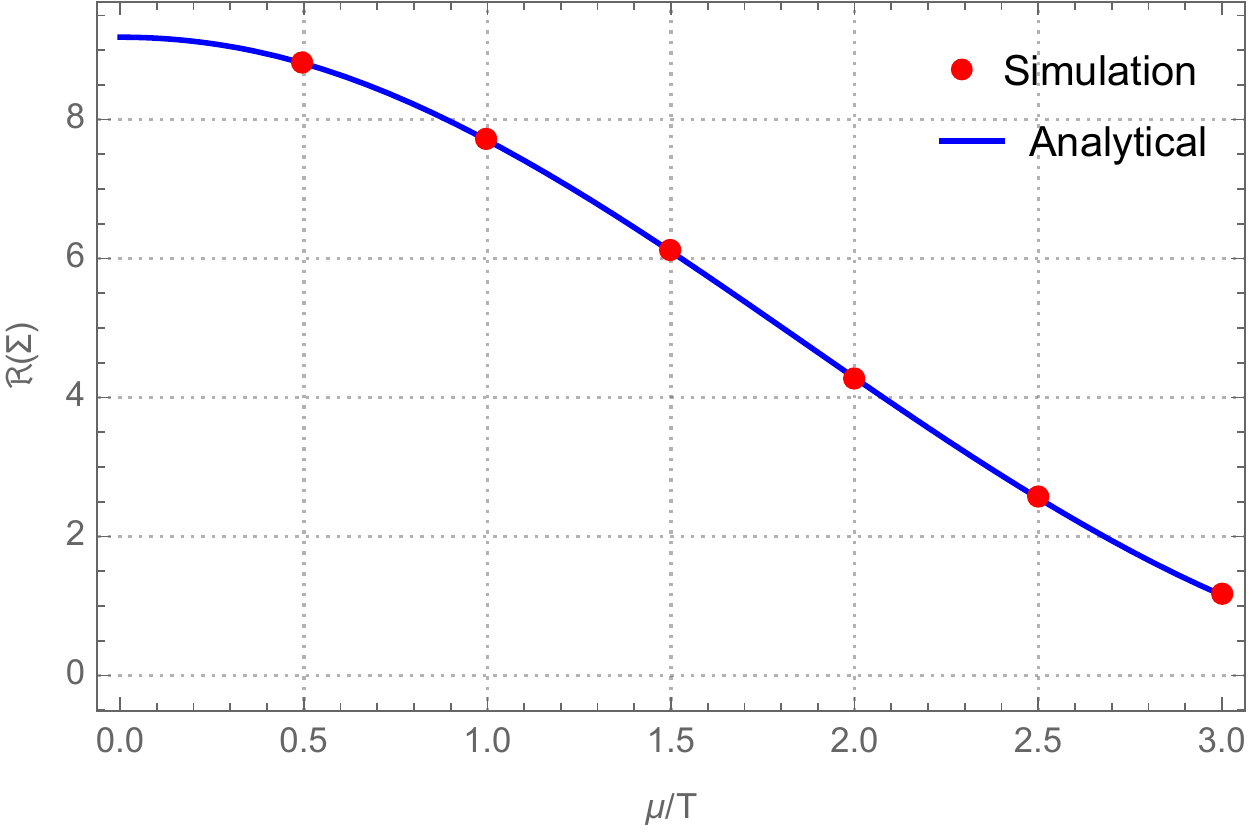}
  }
  \subfloat[$m=1$, $N_f=6$]{
    \label{fig:qcd1_cc_m01_6}
    \includegraphics[height=2.5cm]{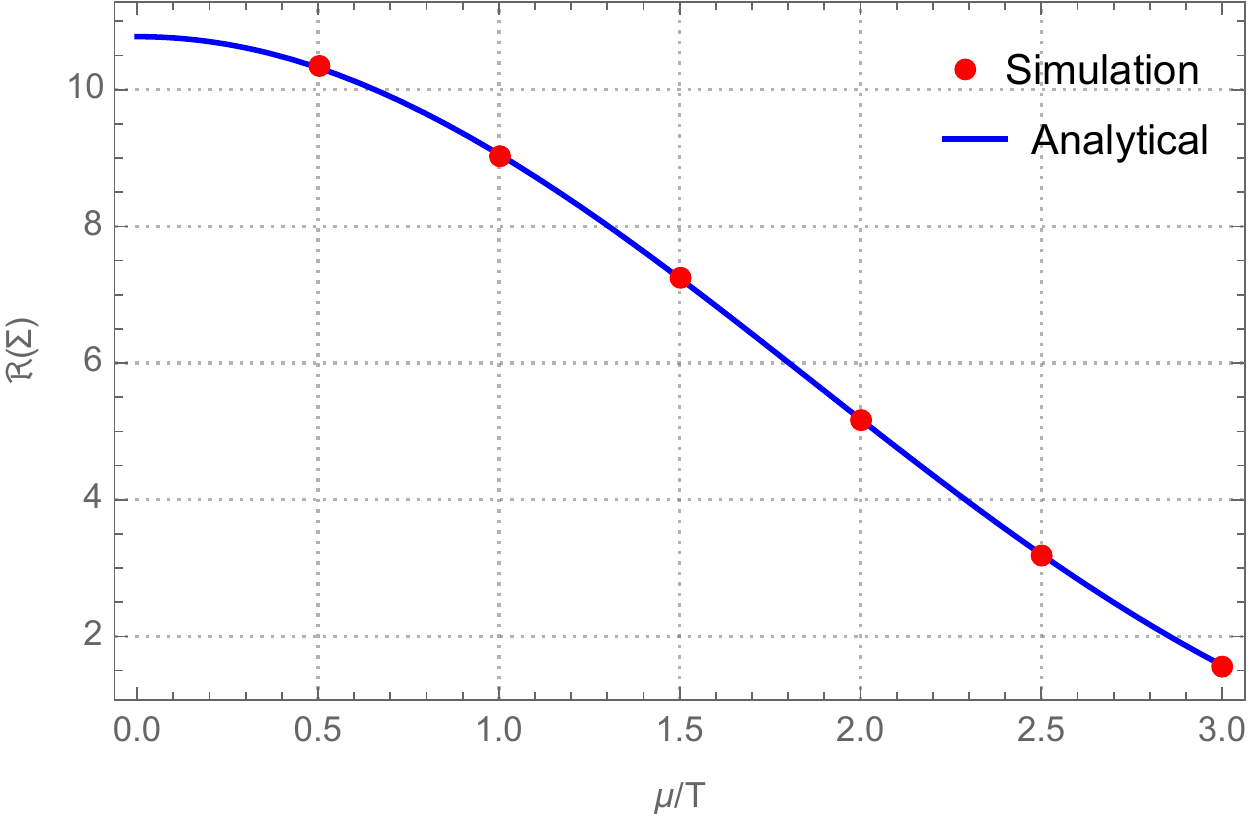}
  }
  \subfloat[$m=1$, $N_f=12$]{
    \label{fig:qcd1_cc_m01_12}
    \includegraphics[height=2.5cm]{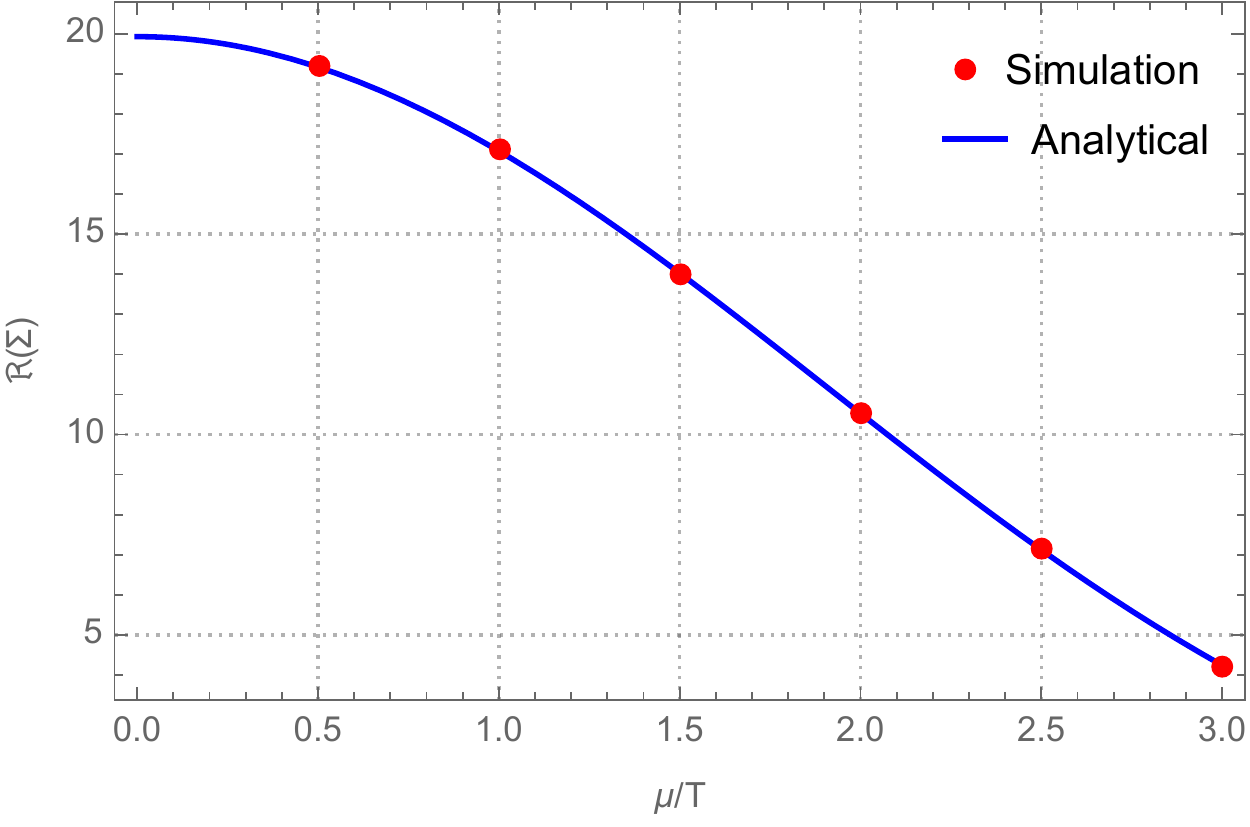}
  }
  \\
  \centering
  \subfloat[$m=1$, $N_f=5$]{
    \label{fig:qcd1_tru_m1_5}
    \includegraphics[height=2.5cm]{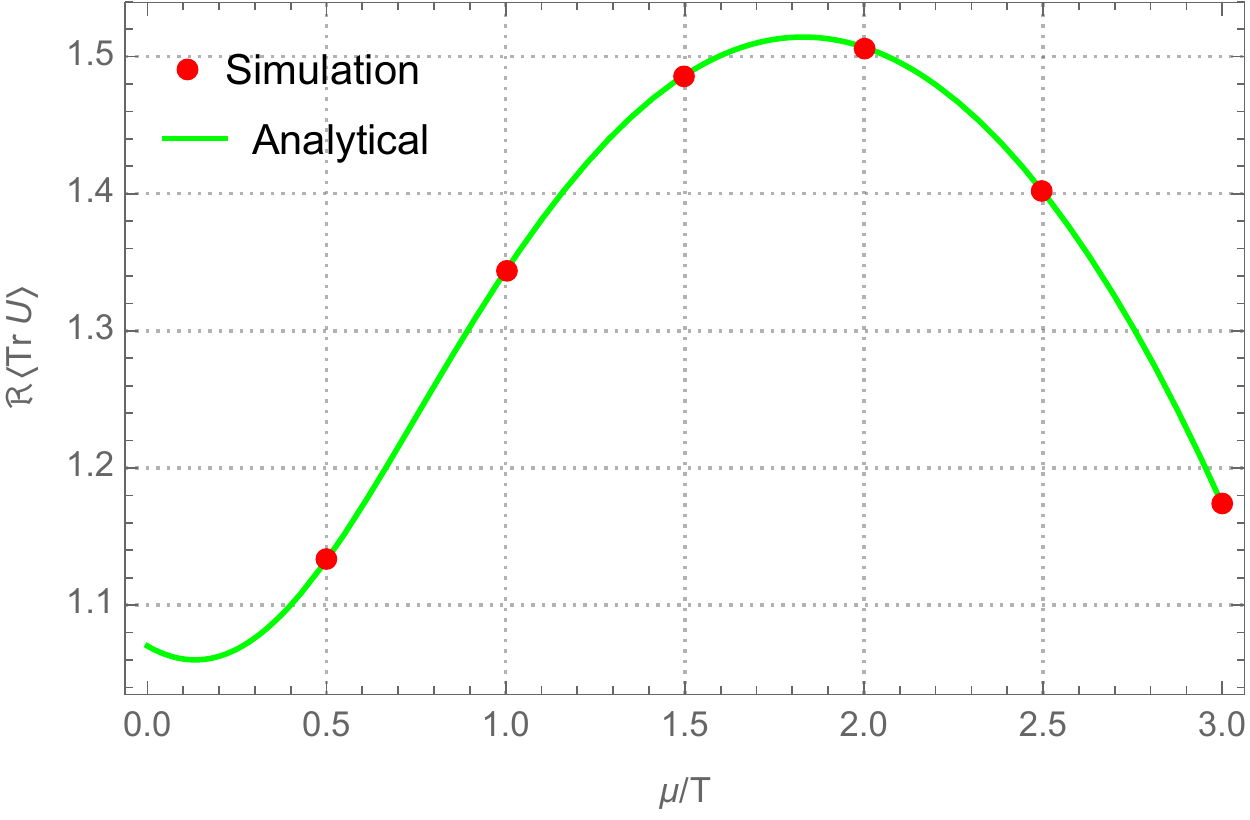}
  }
  \subfloat[$m=1$, $N_f=6$]{
    \label{fig:qcd1_tru_m1_6}
    \includegraphics[height=2.5cm]{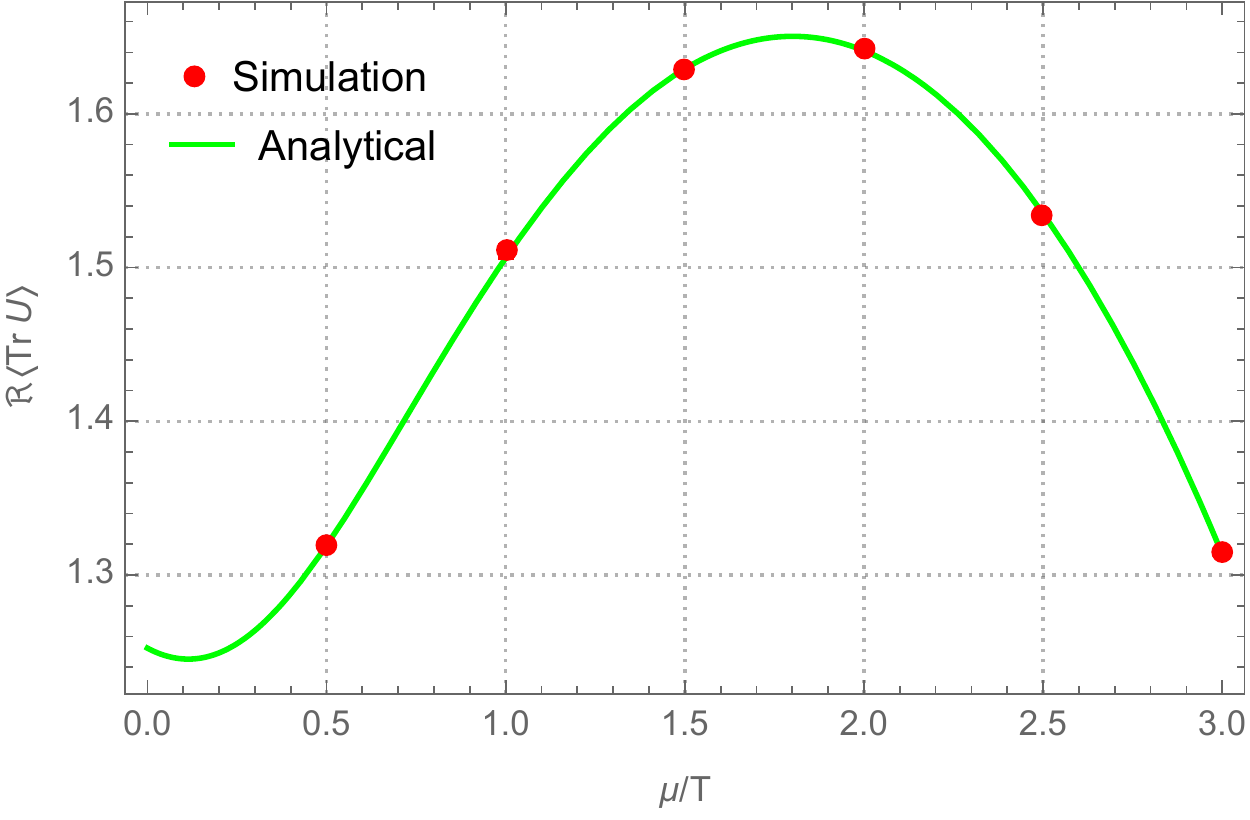}
  }
  \subfloat[$m=1$, $N_f=12$]{
    \label{fig:qcd1_tru_m1_12}
    \includegraphics[height=2.5cm]{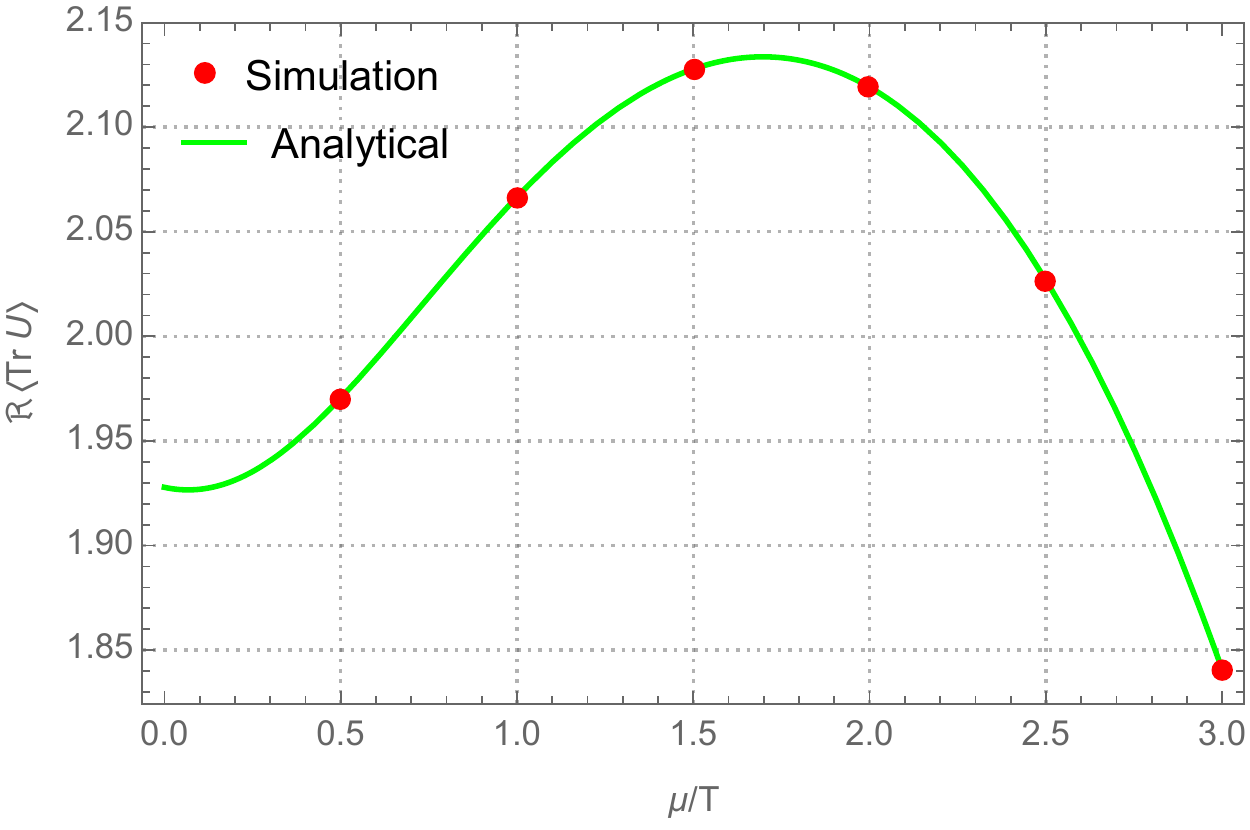}
  }
  \caption{Chiral condensate (blue; 1st and 3rd rows) and Polyakov loop
    (green; 2nd and 4th rows) expectation value for 0+1 QCD at $T=0.5$,
    $m=0.1$ (1st and 2nd rows) and $m=1$ (3rd and 4th rows) for
    $N_f=5,6,12$.  Observables are plotted vs $\mu/T$.}
  \label{fig:qcd1_Nf5.6.12}
\end{figure}

In Figures \ref{fig:qcd1_Nf1.2.3.4} and \ref{fig:qcd1_Nf5.6.12} 
we display results obtained following this procedure for 
the chiral condensate and the Polyakov loop. We cover a range of
values for $N_f$, $m$ and $\mu/T$. 

\begin{figure}[ht]
  \centering
  \subfloat[$N_f=1$]{
    \label{fig:qcd1_r_1}
    \includegraphics[height=2.5cm]{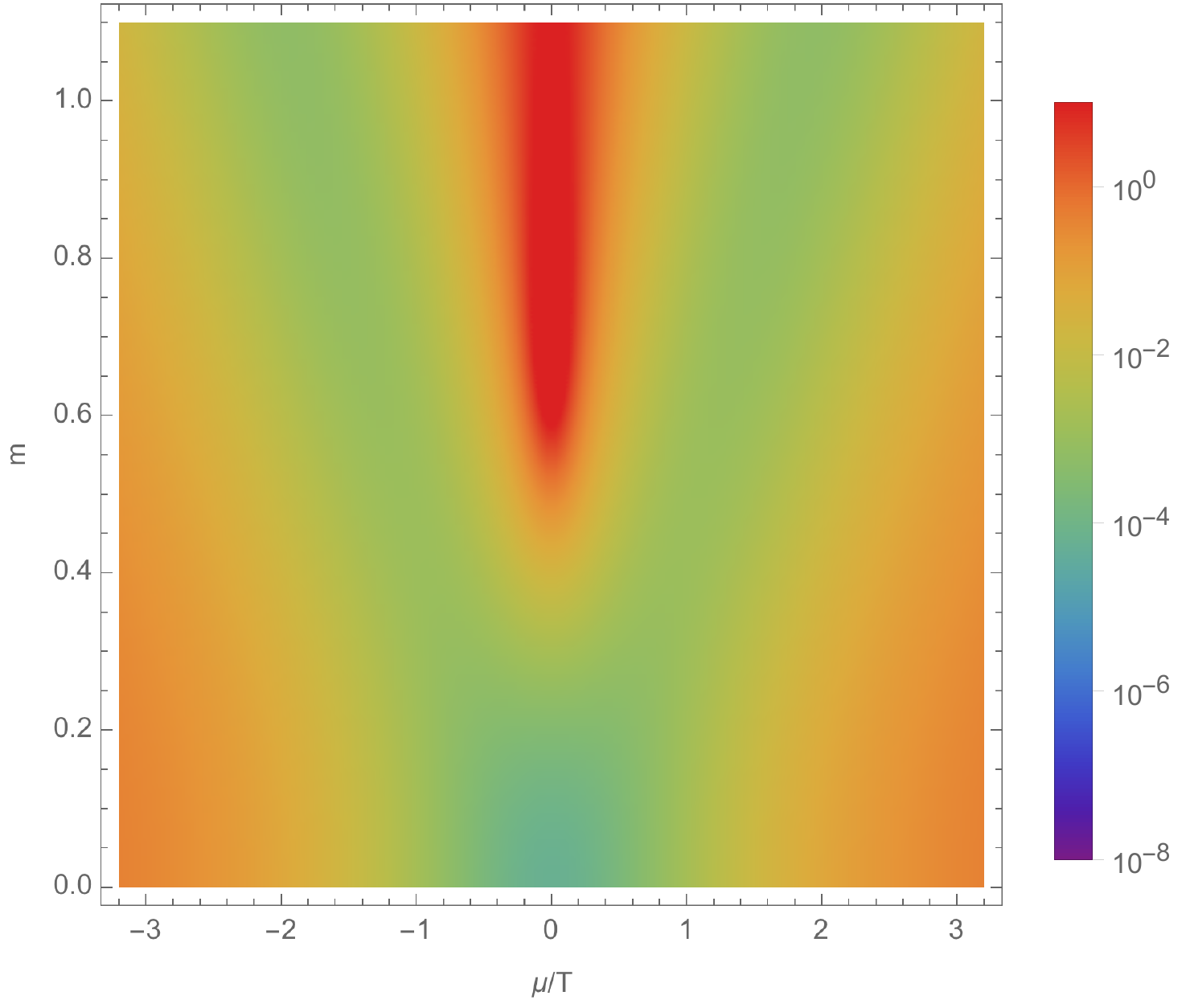}
  }
  \subfloat[$N_f=3$]{
    \label{fig:qcd1_r_3}
    \includegraphics[height=2.5cm]{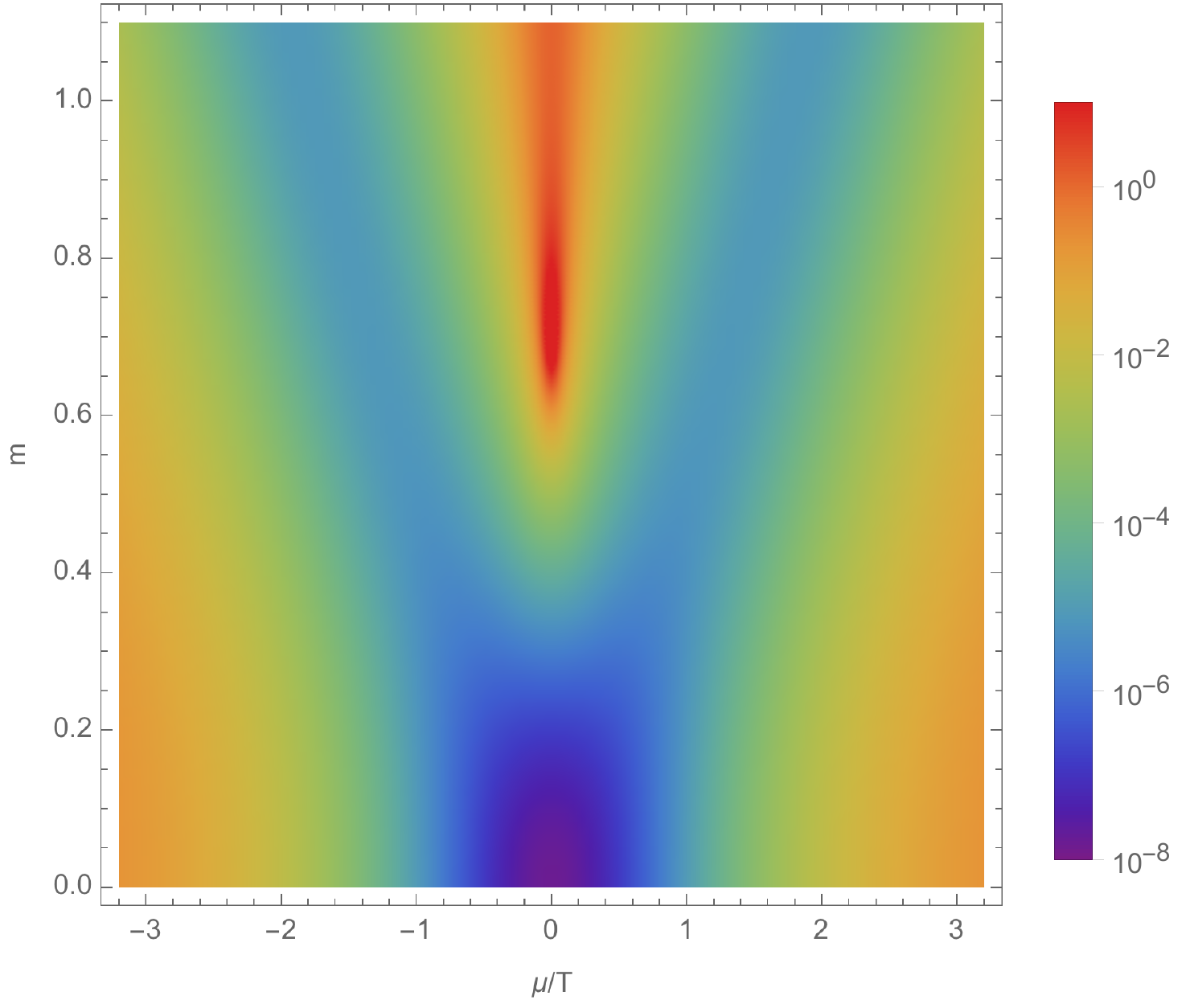}
  }
  \subfloat[$N_f=4$]{
    \label{fig:qcd1_r_4}
    \includegraphics[height=2.5cm]{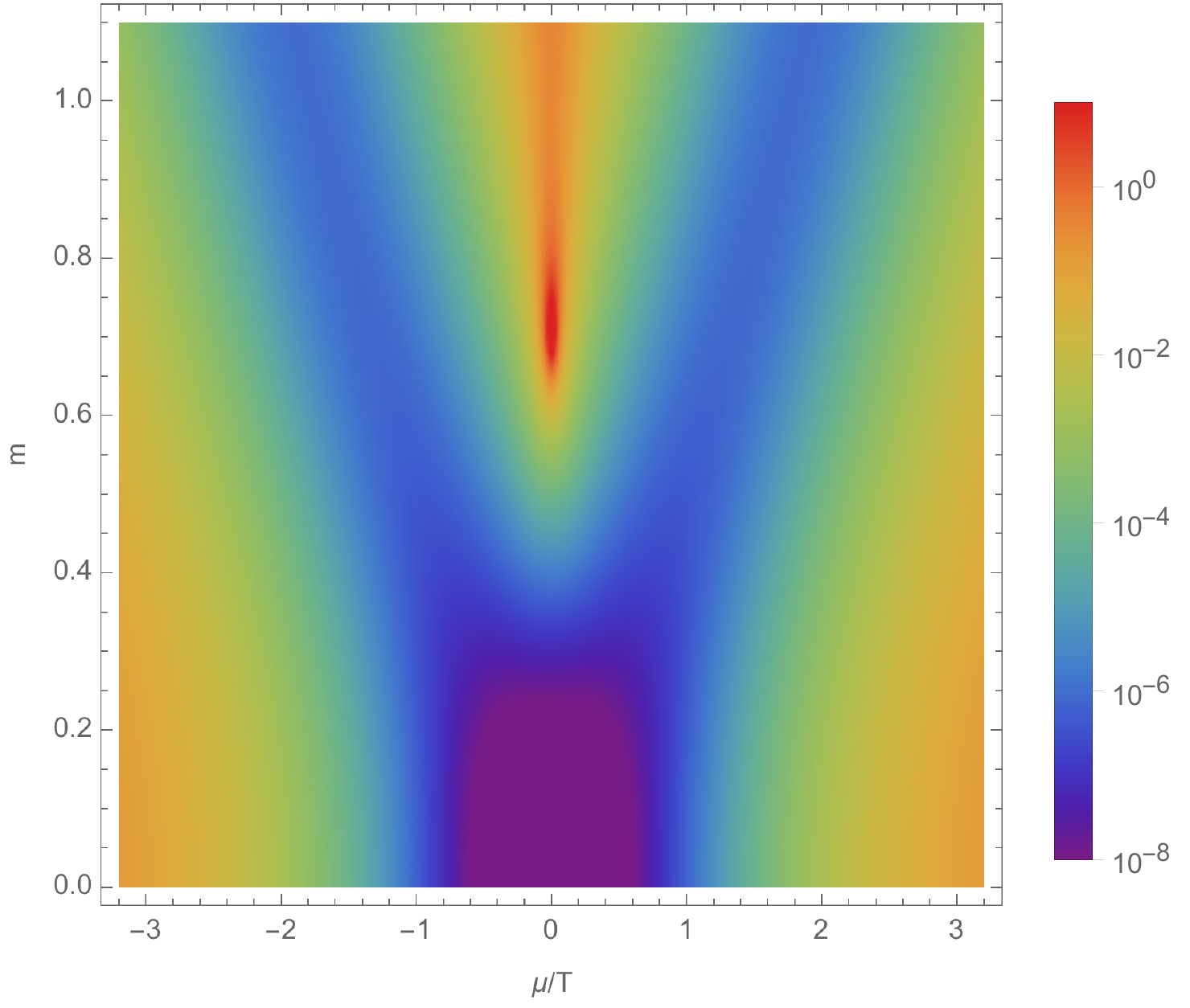}
  }
  \subfloat[$N_f=5$]{
    \label{fig:qcd1_r_5}
    \includegraphics[height=2.5cm]{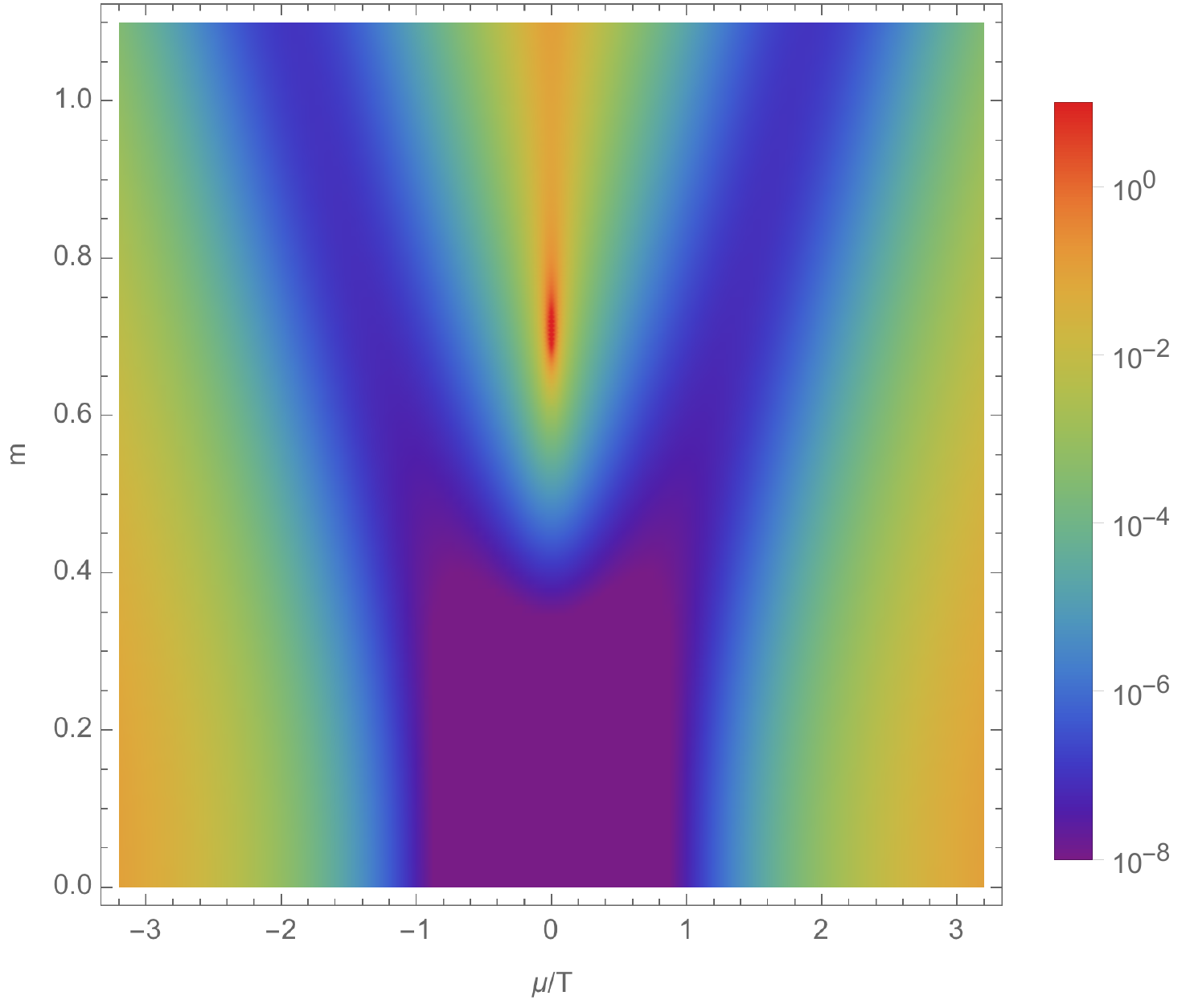}
  }
  \subfloat[$N_f=12$]{
    \label{fig:qcd1_r_12}
    \includegraphics[height=2.5cm]{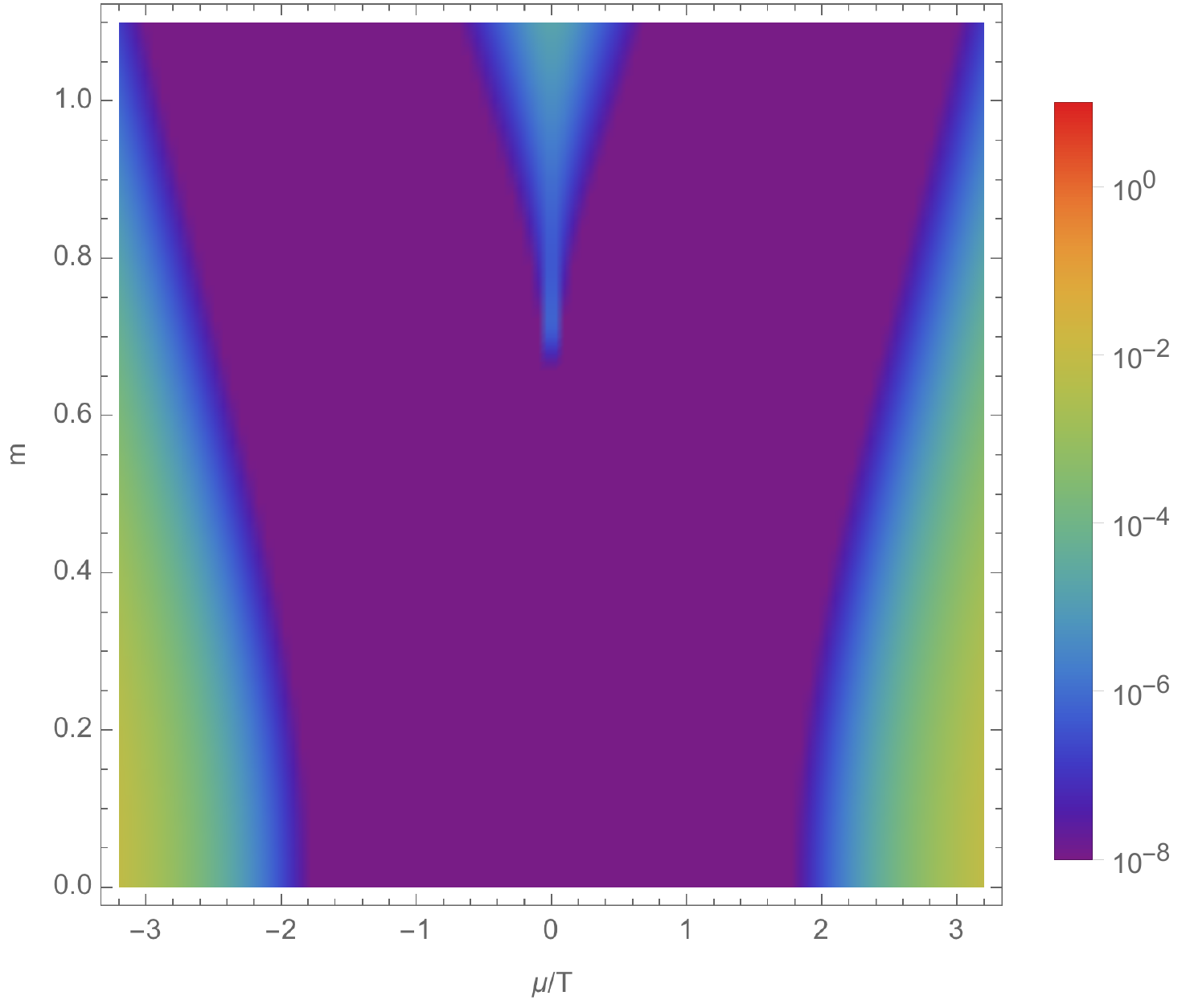}
  }
  \\
  \centering
  \subfloat[$N_f=2$]{
    \label{fig:qcd1_r_2}
    \includegraphics[height=2.5cm]{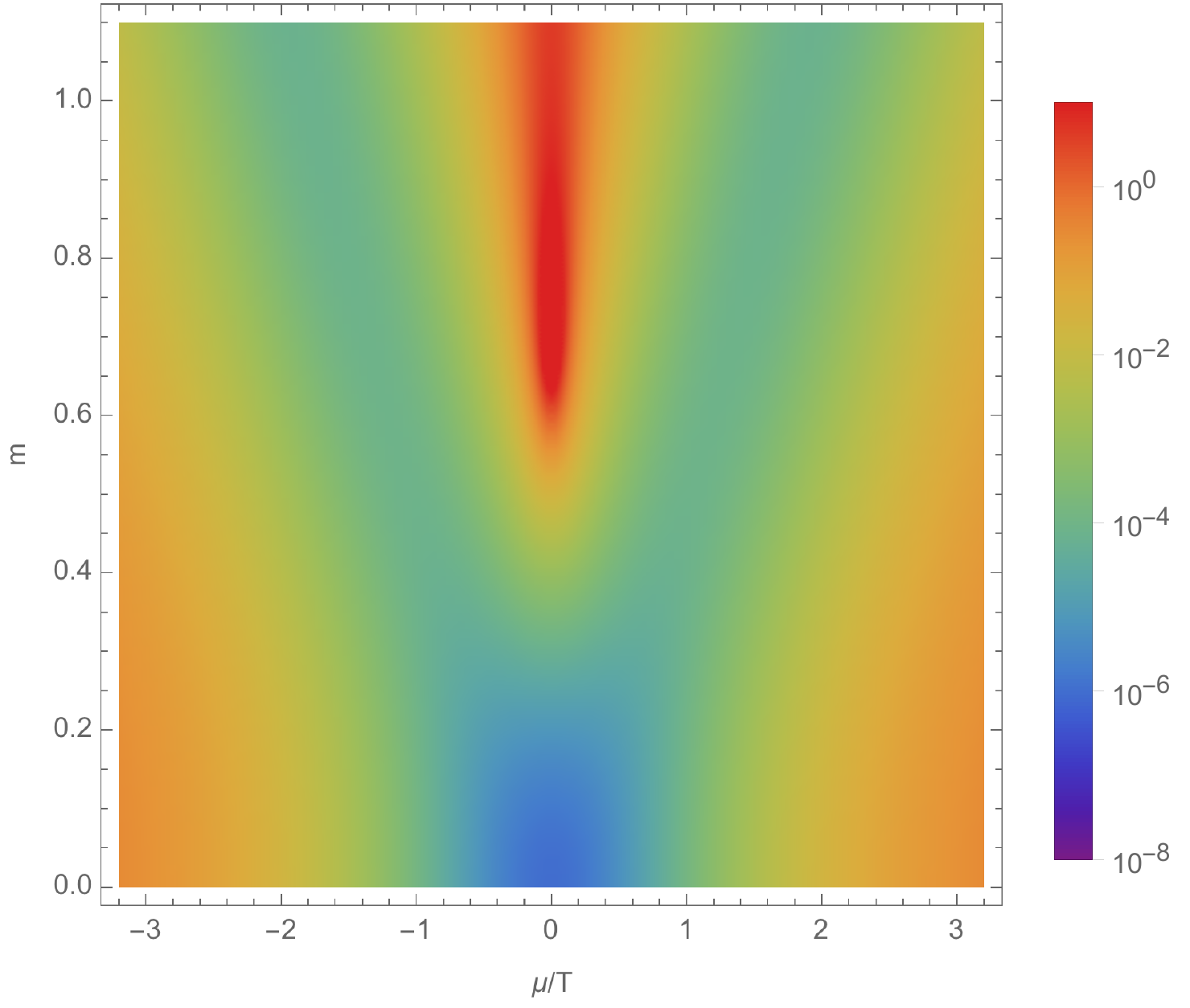}
  }
  \subfloat[$m=1$, only $\mcal{J}_0$]{
    \label{fig:qcd1_tru_m1_2_only0}
    \includegraphics[height=2.5cm]{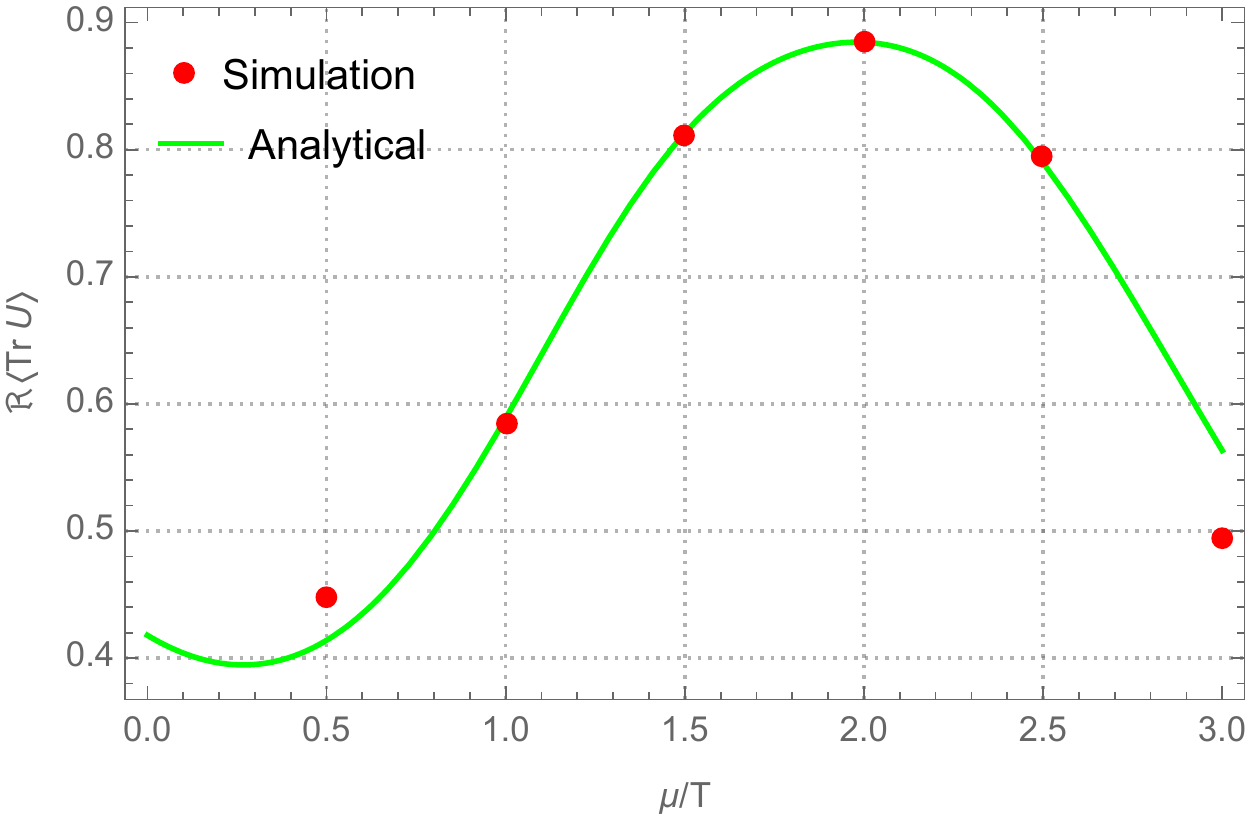}
  }
  \subfloat[$m=1$, all $\mcal{J}_i$]{
    \label{fig:qcd1_tru_m1_2}
    \includegraphics[height=2.5cm]{qcd1_tru_m1_2.pdf}
  }
  \\
  \centering
  \subfloat[$N_f=6$]{
    \label{fig:qcd1_r_6}
    \includegraphics[height=2.5cm]{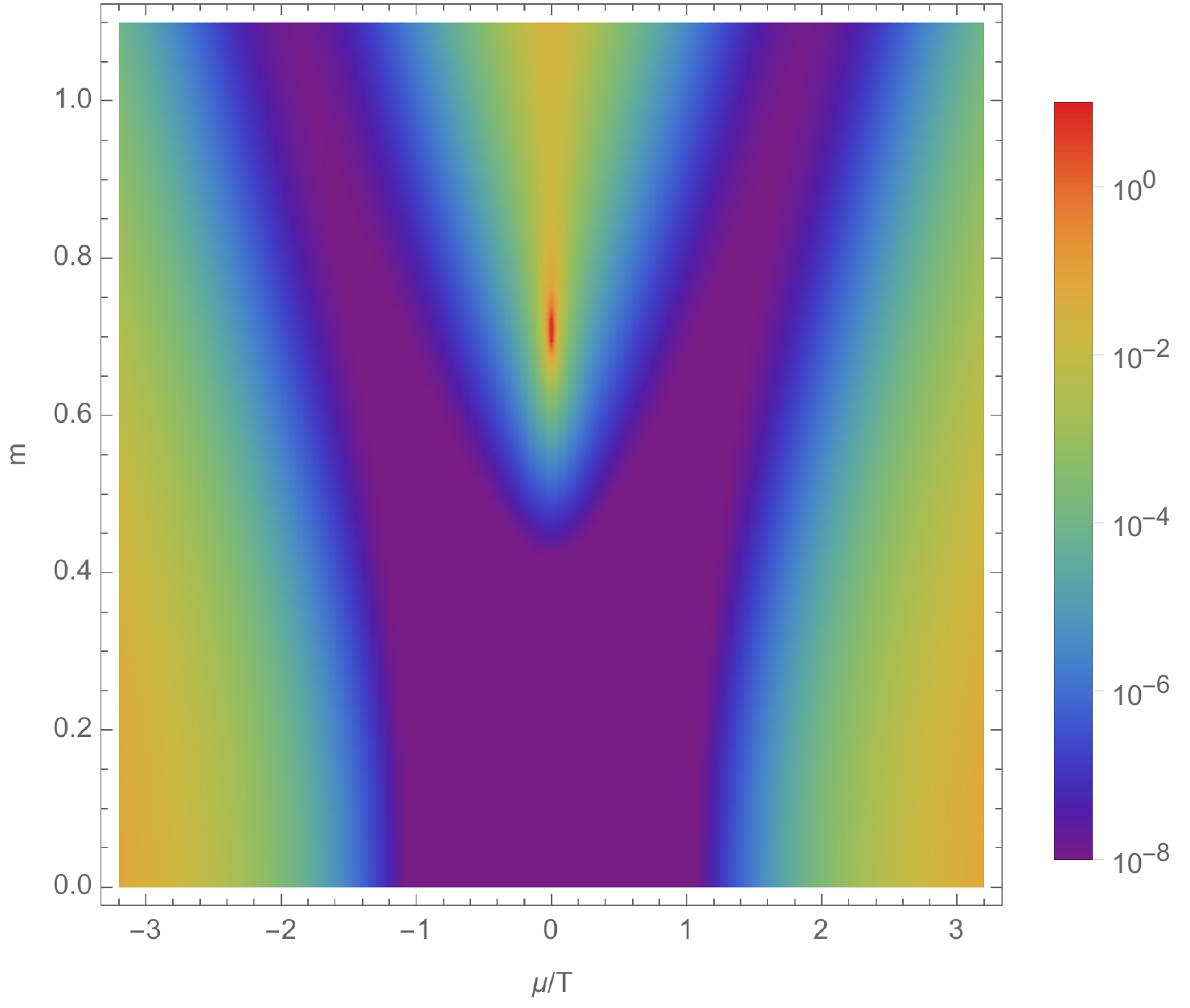}
  }
  \subfloat[$m=0.1$, only $\mcal{J}_0$]{
    \label{fig:qcd1_tru_m01_6_only0}
    \includegraphics[height=2.5cm]{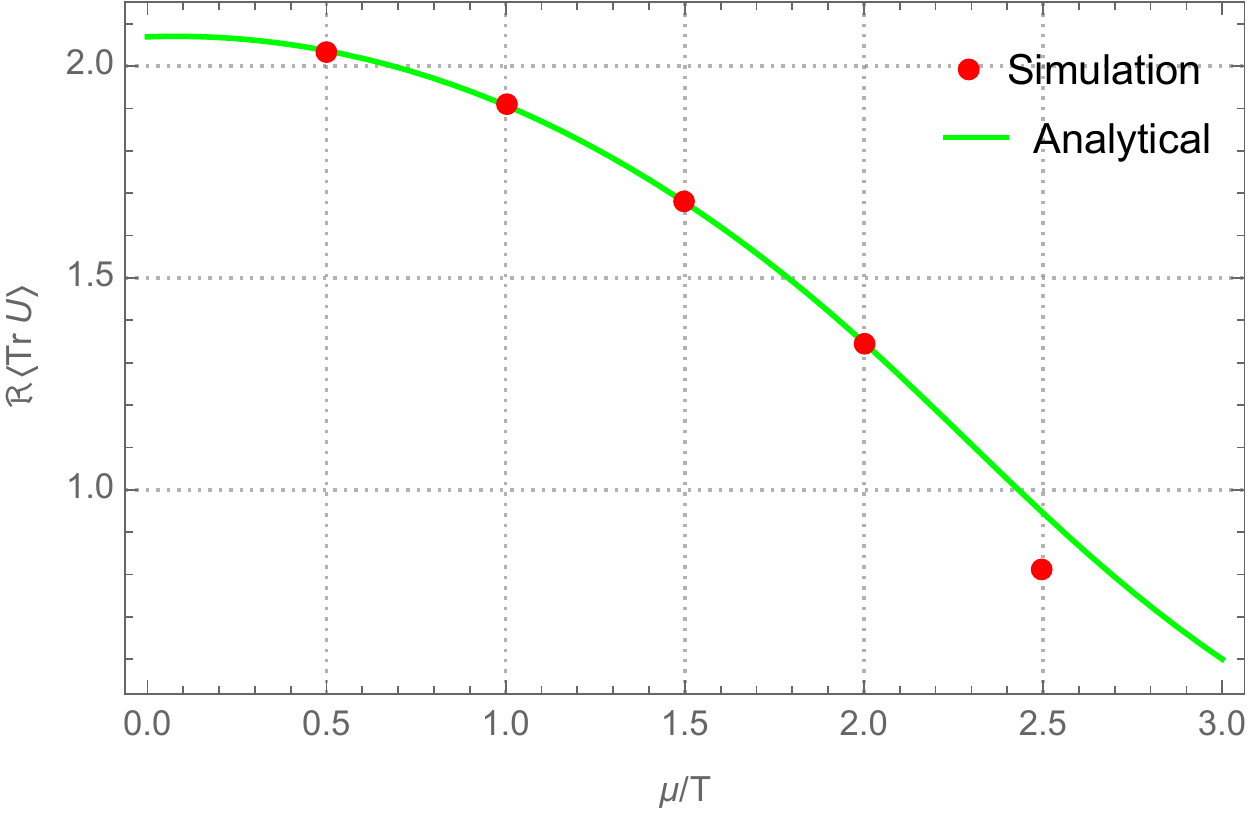}
  }
  \subfloat[$m=0.1$, all $\mcal{J}_i$]{
    \label{fig:qcd1_tru_m01_6}
    \includegraphics[height=2.5cm]{qcd1_tru_m01_6.pdf}
  }
  \caption{First line: $r^{1,2}_0$ as a function of $\mu/T$ (horizontal axis) and $m$
    (vertical axis) for 0+1 QCD at $T=0.5$ and $N_f=1,3,4,5,6,12$.
\newline Second line: $r^{1,2}_0$ as a function of $\mu/T$ and $m$ for
$N_f=2$ and Polyakov loop at $m=1$ integrating only on $\mcal{J}_0$ or
on all thimbles 
\newline Third line: $r^{1,2}_0$ as a function of $\mu/T$ and $m$ for
$N_f=6$ and Polyakov loop at $m=0.1$ integrating only on $\mcal{J}_0$ or
on all thimbles}
  \label{fig:qcd1_r}
\end{figure}

We used this very same flat Monte Carlo prescription to solve 
the Chiral Random Matrix model in \cite{thimbleCRM}, but here there is 
a noticeable difference. In \cite{thimbleCRM} the contribution from
one single thimble was needed. Here, as we said, all the three critical points
belong to the original domain of integration ($SU(3)$) and 
all of them are in principle relevant in the thimble decomposition. 
However, the semiclassical arguments of \ref{sec:semiclassics} 
provide a deeper insight with respect to the actual weight of each 
contribution entering such decomposition. 
Figure \ref{fig:qcd1_r} (first line) depicts 
$r^{1,2}_0$ (defined in (\ref{eq:qcd01_r120})) as a function of 
$\frac{\mu}{T}$ and $m$; by studying this quantity one can predict 
for which values of the parameters $(\frac{\mu}{T},m)$ integration 
only over $\mcal{J}_0$ is expected to capture substantially correct 
results. On the other side (second and third line) figure \ref{fig:qcd1_r}
also shows that there are regions in which taking all the thimbles into
account is compelling: for $m=1$, $N_f=2$ and $m=0.1$, $N_f=6$ the
semiclassical evaluation of $r^{1,2}_0$ predicts that taking only
$\mcal{J}_0$ into account is going to miss the correct result. This is
indeed confirmed by the evaluation of the Polyakov loop. 
We point out that (as
predicted by the semiclassical approximation) at large values of $N_f$
(\eg $N_f=12$) the contribution from the thimble attached to $U_0$
essentially captures the correct results: {\em a fortiori} for even
higher values of $N_f$, {\em the 
single thimble dominance scenario indeed holds true}.

The reader will notice that (in particular in figure
\ref{fig:qcd1_Nf1.2.3.4}) 
simulation results are not shown beyond certain 
values of $\mu/T$ which are dependent on $m$ and $N_f$. At higher 
values of $N_f$, flat Monte Carlo simulations were successful at all values of $\mu/T$. 
This is consistent with the observation that semiclassical estimates 
(which rely on the isotropy of the Hessian spectrum) become exact 
in the limit $N_f\rightarrow\infty$, thus rendering the model easy 
to simulate at high $N_f$ even by flat, crude Monte Carlo. On the
other side, for other 
values of parameters (namely, large $\mu/T$ at small $N_f$) we needed 
to tackle the problem by different means.

\subsection{Simulations by importance sampling}

We now discuss a method to perform importance sampling on
thimbles. For the sake of simplicity, we start in a simplified 
setting, \ie as if only one thimble
contributed. This assumption will be later released to make contact
with the case at hand: in the meantime, this assumption 
allows a simplified notation (for the sake of
notational simplicity we will often omit in the following the
subscript/superscript $\sigma$, \eg in Takagi values: there is no 
need to distinguish since it is assumed that 
only one critical point does matter). 

In the simplified framework of a single thimble contributing, 
the computation of (\ref{eq:basicO}) simply amounts to
\begin{equation}
\langle O\rangle=\frac{\langle O\,e^{i\,\omega}\rangle_\sigma}{\langle
  e^{i\,\omega}\rangle_\sigma}
\label{eq:obs_reweighted}
\end{equation}
where a reweighting with respect to the critical phase is in place and
we introduced the notation
\[
\langle\ldots\rangle_\sigma=\frac{\int\limits_{\mathcal{J}_\sigma}\mathrm{d}^ny\,\ldots\,e^{-S_R}}
{\int\limits_{\mathcal{J}_\sigma}\mathrm{d}^ny\,e^{-S_R}}.
\]
The reader will recognize the expression for $Z_\sigma$ 
(\ref{eq:single_thimble_Z}) in the denominator. 
Making use of the representation (\ref{eq:nNt}), and thus of 
the same notation in which we wrote (\ref{eq:totalZ_Zn}) 
and (\ref{eq:partialZ}), we can now rephrase
\begin{equation}
\langle f \rangle_\sigma=
\frac{1}{Z_\sigma}\int\limits_{\mathcal{J}_\sigma}\mathrm{d}^ny\,f\,e^{-S_R}=\frac{1}{Z_\sigma}\int\mathrm{D}\hat{n}\;
(2\sum_{i=1}^n\lambda_in_i^2) \int\limits_{-\infty}^{+\infty}
\mathrm{d}t\,f(\hat{n},t)\,
e^{-S_{\mathrm{eff}}(\hat{n},t)}=\int\mathrm{D}\hat{n}\,\frac{Z_{\hat{n}}}{Z_\sigma}\,f_{\hat{n}}\label{eq:MC_expect_f}
\end{equation}
in which
\[
f_{\hat{n}}\equiv\frac{1}{Z_{\hat{n}}}
(2\sum_{i=1}^n\lambda_in_i^2) \int\limits_{-\infty}^{+\infty}
\mathrm{d}t\,f(\hat{n},t)\,
e^{-S_{\mathrm{eff}}(\hat{n},t)}
\]
almost looks like a functional integral along a single complete flow
line. (\ref{eq:MC_expect_f}) can be put at work in the computation of
(\ref{eq:obs_reweighted}) (with $f=O\,e^{i\,\omega}$ in the numerator 
and $f=e^{i\,\omega}$ in the denominator). 
(\ref{eq:MC_expect_f}) is nothing but the average of the
$f_{\hat{n}}$, {\em i.e.} the average of the contributions that a 
given observable takes from complete flow lines, where the weight 
$\frac{Z_{\hat{n}}}{Z_\sigma}$ represents the fraction of the
partition function which is provided by a single complete flow line.
$\frac{Z_{\hat{n}}}{Z_\sigma}$ provides a natural setting
for importance sampling: {\em directions} $\hat{n}$ have to be
extracted according to the probability 
$P(\hat{n})=\frac{Z_{\hat{n}}}{Z_\sigma}$. 

We proceed as follow. 
In our Markov chain we start from the current configuration (which is
associated to a direction $\hat{n}$) and we propose a new one 
(associated to a direction $\hat{n}'$). $\hat{n}'$ is identical to $\hat{n}$ 
apart from two randomly chosen components, 
say $(n_i,n_j)$ with $i\neq j$. We define $C$ by
\[
C\equiv n_i^2+n_j^2=\mathcal{R}-\sum_{k\neq i,j}n_k^2
\]
which is fixed by the normalization
$\left|\vec{n}\right|=\sqrt{\mathcal{R}}$ and by the values of all
$\{n_k\}_{k\neq i,j}$. There is a coordinate system in which we can 
now parametrize the current values of $(n_i,n_j)$ by
\[
n_i=\sqrt{C}\,\cos\phi \;\;\;\;\; n_j=\sqrt{C}\,\sin\phi
\]
with $\phi\in[0,2\pi)$. Our evolution step now amounts to change 
$\phi \rightarrow \phi'$, which results in $(n_i,n_j) \rightarrow 
(n_i',n_j')$, while for all the other components (${k\neq i,j}$) 
$n_k = n_k'$. $\phi'-\phi$ is extracted flat in a given (tunable)
range. 
We finally accept the proposed configuration with the
standard Metropolis test
\begin{equation}
P_{\mathrm{acc}}\left(\hat{n}'\bigr|\hat{n}\right)=
\min\left\{1,\frac{Z_{\hat{n}'}}{Z_{\hat{n}}}.
\right\}\label{eq:MC_flatmetroacc}
\end{equation}

We actually have a more efficient Monte Carlo (which has been
preliminary described in \cite{LAT16} and will be further discussed
elsewhere \cite{MCsoon}): we notice that this is not applicable here,
due to the full degeneracy of the gaussian spectrum of the theory at
hand. 

In our case three contributions should be in principle taken into
account. Actually, due to the symmetry of Section \ref{sec:symmYuya},
only two distinct contributions are in place and
(\ref{eq:obs_reweighted})
now reads
\begin{equation}
\langle O\rangle=\frac
{n_0 \, e^{-i\,S_{I 0}} \, Z_0 \, \langle O\,e^{i\,\omega}\rangle_0 \, + \, n_{12}
  \, e^{-i\,S_{I 12}} \, Z_{12} \, \langle O\,e^{i\,\omega}\rangle_{12}}
{n_0 \, e^{-i\,S_{I 0}} \, Z_0 \, \langle e^{i\,\omega}\rangle_0 \, + \, n_{12}
  \, e^{-i\,S_{I 12}} \, Z_{12} \, \langle e^{i\,\omega}\rangle_{12}}
\label{eq:obs_reweighted_2}
\end{equation}

\noindent
in which notations should be evident (\eg $Z_0$ is associated to thimble 
$\mathcal{J}_0$ and $Z_{12}$ is the average of the contributions
associated to thimbles $\mathcal{J}_1$ and $\mathcal{J}_2$). In order
to proceed we now put at work a recipe that we put forward in
\cite{thimbleCRM}. Equation (\ref{eq:obs_reweighted_2}) can be rewritten 

\begin{equation}
\langle O\rangle=\frac
{\langle O\,e^{i\,\omega}\rangle_0 \, + \, \alpha
\langle O\,e^{i\,\omega}\rangle_{12}}
{\langle e^{i\,\omega}\rangle_0 \, + \, \alpha
\langle e^{i\,\omega}\rangle_{12}}
\label{eq:obs_reweighted_3}
\end{equation}
where we defined 
\begin{equation}
\alpha \equiv \frac
{n_{12} \, e^{-i\,S_{I 12}} \, Z_{12}}
{n_0 \, e^{-i\,S_{I 0}} \, Z_0}
\label{eq:DEFalpha}
\end{equation}

The idea is now to determine the value of $\alpha$ taking a given
observable as a {\em normalization point}: all the other observables of
the theory can then be computed using this input. This can indeed be
done. As an example, at $N_f=1, m=0.1, \mu/T=0.5$ we get
$\alpha=0.2686(13)$ and $\alpha=0.2682(8)$ (taking different
observables as normalization points). 
\begin{figure}[t]
\begin{center}
\includegraphics[height=11cm,clip=true]{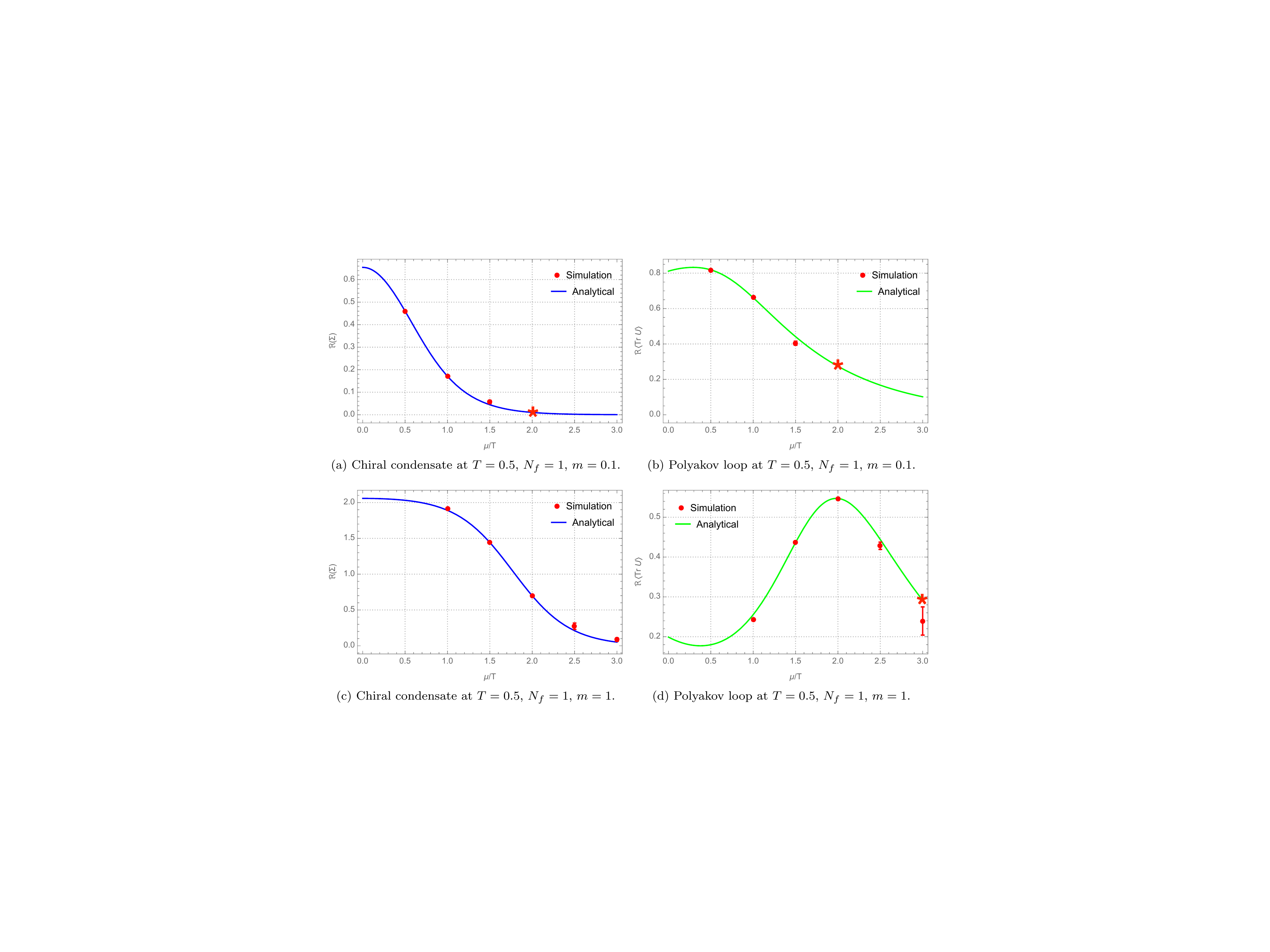}
\caption{Extending the range of computations of chiral condensate and
  Polyakov loop to values of $\mu/T$ for which flat, crude Monte Carlo
  was failing. New points obtained by importance sampling and making use of
  (\ref{eq:obs_reweighted_3}) are drawn as stars: errorbars are
  negligible with respect to the symbols size.}
\label{fig:newResults}
\end{center}
\end{figure}
Figure \ref{fig:newResults} confirms the effectiveness of the
procedure. For $N_f=1, m=0.1$ the point $\mu/T=2.0$ was completely out
of reach for flat, crude Monte Carlo, while results are successfully
computed with the improve metod (first row). Notice that for these
values of parameters the tiny value of the chiral condensate (of order
$10^{-2}$) results from a delicate cancelation of the contribution 
coming from the different thimbles. While this could be seen as a 
sign problem coming back, one should notice that it is a numerical 
accident occurring for a given observable at a given value of
parameters. On the other side, it has nothing to do with the original
sign problem (nor \eg with the residual phase). Figure
\ref{fig:newResults} also shows (second row) how errorbars can be cut
down in the case of Polyakov loop for $N_f=1, m=1, \mu/T=3.0$.

\section{Conclusions and prospects}

We showed that QCD in 0+1 dimensions can be effectively computed in
thimble regularization. The thimbles attached to three critical points
are in principle relevant, but a symmetry relates the results coming
from two of them. Importance sampling in the space of steepest ascents
enabled us to successfully compute observables in regions where a
flat, crude Monte Carlo was failing. Nevertheless, the latter has
performed reasonably well on a quite extended region of parameter 
space.

The sign problem which is place in 0+1 dim QCD is a genuine one,
stemming from the Dirac determinant. On the other side, the theory at
hand does not show the subtleties of gauge symmetry: the task in front
of us is now to proceed to tackle gauge theories.

\section*{Acknowledgments}
\par\noindent
We acknowledge support from I.N.F.N. 
under the research project {\sl i.s. QCDLAT}.
This work is dedicated to Franz Scorzato. 
%\newpage

\end{document}